\documentstyle[aps,prb,multicol,epsf]{revtex}  
\catcode`\@=11       
\def\listzigurename{List of Zigures}
\def\zigurename{Zigure}
\def\listofzigures{\section*{\listzigurename
    \@mkboth{\uppercase{\listzigurename}}{\uppercase{\listzigurename}}}%
  \@starttoc{lof}}

\def\l@zigure{\@dottedtocline{1}{1.5em}{2.3em}}

\let\l@table\l@zigure

\newcounter{zigure}
\def\thezigure{\@arabic\c@zigure}

\def\fps@zigure{tbp}
\def\ftype@zigure{1}
\def\ext@zigure{lof}
\def\fnum@zigure{\zigurename~\thezigure}
\def\zigure{\@float{zigure}}
\let\endzigure\end@float
\@namedef{zigure*}{\@dblfloat{zigure}}
\@namedef{endzigure*}{\end@dblfloat}
\catcode`\@=12
\begin{document}                
\title{On the low energy properties of fermions with singular interactions}
\author{B. L. Altshuler}
\address{Department of Physics, MIT, Cambridge, MA 02139}
\author{L. B. Ioffe}  
\address{Department of Physics, Rutgers University, Piscataway, NJ 08855 \\
and Landau Institute for Theoretical Physics, Moscow, Russia}
\author{A. J. Millis}  
\address{AT\&T Bell Laboratories, Murray Hill, NJ 07974}
\maketitle
\begin{abstract}                
We calculate the fermion Green function and particle-hole susceptibilities for
a degenerate two-dimensional fermion system with a singular gauge interaction.
We show that this is a strong coupling problem, with no small parameter
other than the fermion spin degeneracy, N.
We consider two interactions, one arising in the context of the $t-J$
model and the other in the theory of half-filled Landau level.
For the fermion self energy we show in contrast to previous claims 
that the qualitative behavior found in the
leading order of perturbation theory is preserved to all orders in the
interaction.  The susceptibility $\chi_Q$ at a general wavevector $\bf{Q} \neq 2\bf{p_F}$
retains the fermi-liquid form. However
the $2p_F$ susceptibility   $\chi_{2p_F}$ either
diverges as $T\rightarrow 0$ or remains finite but with nonanalytic
wavevector, frequency and temperature dependence.
We express our results in the language of recently discussed scaling theories,
give the fixed-point action, and show that at this fixed point the fermion-gauge-field
interaction is marginal in $d=2$, but irrelevant at low energies in $d \ge 2$.
\end{abstract}

\begin{multicols}{2}
\section{Introduction}

The problem of fermions in two dimensions interacting with a singular gauge
interaction has arisen recently in two physical contexts.
One is the ``gauge theory'' approach \cite{ioffelarkin,bza} to the $t-J$ model 
which has been argued \cite{anderson,rice} to contain the essential
physics of high $T_c$ superconductors.
The other is the theory of the half-filled Landau level
\cite{halperinleeread,kalmeyerzhang}.
In both cases one is led to the theoretical problem of a degenerate Fermi gas
interacting with a gauge field characterized by the propagator $D(\omega,k)\sim
(\frac{|\omega|}{k} + |k|^{1+x})^{-1}$.
This notation is conventional; $x=1$ in the $t-J$ model case
\cite{ioffelarkin} and, because of 
the unscreened Coulomb interaction, $x=0$ in the $\nu=1/2$ case considered by
previous authors \cite{halperinleeread,kalmeyerzhang}. 
If the Coulomb interaction in two dimensions were screened, e.g. by a metallic
gate, the model with $x=1$ would apply even to the $\nu=1/2$ case.
The three dimensional version of this model with $x=1$ was shown by Reizer 
\cite{reizer} to describe electrons in metals interacting  magnetically via
a current-current interaction.
The highly singular behavior of the gauge propagator at small 
$\omega, k$ complicates the analysis of the theory and has led to conflicting
claims in the literature.
In this paper we present what we believe is a correct treatment
of the low energy properties of the theory.

We study  fermions interacting with a gauge field $\bf a$,  and also with
each other via a short range interaction $W$.
We assume the simplest form of the interaction with the gauge field
allowed by gauge invariance \cite{ioffelarkin}:
\begin{eqnarray}
H=&&\sum_{p,\sigma}  c^{\dagger}_{p\sigma} \epsilon(p) c_{p\sigma} +
\sum_{p,k,\sigma}  c^{\dagger}_{p+k/2,\sigma} {\bf a_k v}(p) c_{p-k/2,\sigma}
	\nonumber \\	
+&& \sum_{p_i} W c^\dagger_{p_1,\alpha} c_{p_2,\alpha} c^\dagger_{p_3,\sigma}
	c_{p_4,\sigma} \delta(\sum p_i)
+ \frac{1}{4g_0^2} f_{\mu \nu}^2
\label{H}
\end{eqnarray}
where we omitted higher order terms in the gauge field $\bf a$ which lead to 
less infra-red singular effects.
Here, as usual, $f_{\mu \nu}= \partial_\mu a_\nu - \partial_\nu a_\mu$, $g_0$
is the bare fermion-gauge field interaction constant, $\sigma=1..N$ is a 
spin index and ${\bf v}= \frac{\partial \epsilon}{\partial {\bf p}}$. 
The $f_{\mu \nu}^2$ term comes from integrating out high energy processes.
In the $t-J$ model the spin degeneracy $N=2$; however, it will be convenient
to consider general values of $N$ because the limits $N \rightarrow 0 $ and
$N \rightarrow \infty$ are solvable.  Indeed, as we shall see N is the only expansion paramter of the model.

In the $t-J$ model the fermion operators $c^\dagger_{p,\sigma}$ create
``spinons'' which are chargeless, spin $1/2$ quanta.
Because the spinons have no charge, there is no long-range Coulomb term in $W$.
In this representation the charge is carried by different, spinless quanta 
which obey bose statistics (``holons''); we shall not consider their
properties in this paper.
The Hamiltonian (\ref{H}) describes the magnetic properties of the
``spin-liquid'' state of $t-J$ model.
For a more detailed discussion of the physical situations to which this model
may apply, see, e.g., the review paper by Lee \cite{lee-review}.

In the $\nu=1/2$ case the spin degeneracy $N=1$ and one must add to the
Hamiltonian (\ref{H}) additional terms containing the Coulomb interaction and
Chern-Simons term.
This changes the $k^2$ term in gauge propagator to $|k|$; the effects of tthis change will be discussed below in Section IV.
Other authors \cite{nayakwilczek} have found it convenient to consider a 
continously varying
exponent $|k|^{1+x}$; we find that the behavior for all $x>0$ is the same as
that for $x=1$ except for  minor changes in exponents.  The $x=0$ case is exceptional
because there a controlled expansion for the infrared behavior exists for
any N.
We give results for general $x>0$ in Section V.
For the rest of this Section we explicitly consider only the ``spin liquid''
case.

Treating the fermion-gauge field interaction in (\ref{H}) by  perturbation
theory leads immediately to two effects.
Dressing the gauge propagator by a particle-hole bubble leads to the propagator
\begin{equation}
D(\omega,k) = \frac{1}{\frac{Np_0|\omega|}{2\pi |k|} +\frac{1}{N^{1/2}g^2} k^2}
\label{D}
\end{equation}
Here the first term in the denominator is due to Landau damping of the gauge
field, and $p_0$ is the curvature of the Fermi surface at the point where the normal to the fermi surface is perpendicular to $\bf{k}$.
The second  term in the denominator has contributions from the 
$f_{\mu \eta}^2$ term in the effective action and from the fermion 
diamagnetic susceptibility; 
in this term we have redefined the interaction constant $g^2$ so that the 
characteristic energy scale remains
finite in the limits $N\rightarrow \infty$ and $N\rightarrow 0$ which we 
consider below.

Using the gauge field propagator to calculate the fermion self energy $\Sigma$
in the first order of the perturbation theory one finds \cite{lee}
\begin{equation}
\Sigma^{(1)}(\epsilon)=-i \left| \frac{\omega_0}{\epsilon} \right|^{1/3}
        \epsilon
\label{Sigma^(1)}
\end{equation}
where the energy scale $\omega_0$ is defined in terms of $g$, $p_0$ and the fermi
velocity $v_F$ via
\begin{equation}
\omega_0= \left(\frac{1}{2\sqrt{3}} \right)^3  
	\frac{2 v_F^3 g^4}{\pi^2 p_0}
\label{omega_0}
\end{equation}

For high-$T_c$ materials ($N=2$) $g^2$ was estimated to be $6\sqrt{2}\pi m$
(where $m$ is the fermion mass).
This leads to $\omega_0 \sim 500 \;K$ if the fermion bandwidth is of the order
of $2J$.

The dramatic effects found in the leading order of perturbation theory lead
one to question whether the perturbation theory makes sense.
Several different treatments have appeared 
\cite{nayakwilczek,ioffelidsky,houghton,ganwong,khveschenkostamp}.
The appearance of $N$ in the denominator of the gauge field propagator
(\ref{D}) suggests that the theory should have a tractable $N \rightarrow
\infty$ limit and that a $1/N$ expansion about this limit is well behaved.
The $N \rightarrow \infty$ limit and the leading $1/N$ corrections to the
fermion propagator have been studied by Ioffe and Larkin \cite{ioffelarkin}, 
by Reizer \cite{reizer} and by Lee \cite{lee} but the higher order corrections
and the issue of convergence of the expansion have not to our knowledge been 
previously examined. 
We present this analysis in Section II of this paper.
We find that the $1/N$ expansion is indeed well defined and the leading order
results are qualitatively correct for all physical quantities {\em except} 
the $2p_F$ susceptibilities, which acquire additional non-analytic power law
dependences with exponents which vanish as $N \rightarrow \infty$.

In order to explain the idea of the analysis we need to introduce some 
notation and establish typical values of momenta and energies involved in 
virtual processes. 
The typical momentum $k_\omega$ transferred in a low energy process 
affecting a fermion with energy $\omega$ and momentum $p$ is found from  
the gauge field propagator, eq. (\ref{D}), to be
\begin{equation}
k_\omega = N^{1/2} \left( \frac{p_0 g^2 \omega}{\pi} \right)^{1/3}.
\label{k_omega}
\end{equation}
It is convenient to  choose Cartesian coordinates in  
momentum space so that $k_{\perp}$ is the change of the momentum of the 
fermion along the Fermi surface (i.e. perpendicular to $\bf{p}$) and 
$k_{\parallel}$ perpendicular to the fermi surface (i.e. along $\bf{p}$).
At low energies $k_{\parallel} \sim |\omega - \Sigma(\omega)|/v_F$ becomes
much less than  $k_{\perp}$ which is determined by the gauge field propagator
(\ref{D}), i.e. $k_{\perp} \sim k_\omega$. 

Qualitatively the small value of the higher order corrections at large $N$ can
be attributed to a comparatively large typical momentum transfer 
(\ref{k_omega}) in this limit as follows.
In a typical virtual process fermion probes only a small patch of the Fermi
surface of the order of $k_\omega$.
The curvature of this piece of the Fermi surface is important if the change 
in the fermion energy induced in such a virtual process 
($v_F k_{\perp}^2/(2p_0)$) 
is large compared with the imaginary part of its self energy 
(\ref{Sigma^(1)}). 
Comparing the two we find
\begin{equation}
\frac{v_F k_{\epsilon}^2}{2p_0 \Sigma^{(1)}(\epsilon)} \sim N
\label{curvature}
\end{equation}

Thus, in the limit of large $N$ the curvature of the Fermi surface becomes
important and we expect that the scattering becomes essentially two
dimensional.
In this case the usual phase space arguments \cite{migdal} show that all crossing
diagrams are small in $1/N$, so that a $1/N$ expansion is possible.

An alternative solvable limit, namely $N \rightarrow 0$, was pointed out by
Ioffe, Lidsky and Altshuler \cite{ioffelidsky}.
In this limit the curvature of the Fermi surface
becomes unimportant and the terms proportional to $k_\perp^2$ in the
denominators of the fermion Green function are negligible.
When these terms are dropped the Green function does not depend on $k_\perp$,
which enters only via the propagator of the gauge field (\ref{D}).
Thus, in any diagram one can integrate independently all the gauge field
propagators over $k_\perp$.
The gauge field propagator becomes
\begin{equation}
D^{(1D)}(\omega) = \int D(\omega,k) (dk_\perp) = \frac{\tilde{g}}{v_F
|\omega|^{1/3}} 
\label{D^(1D)}
\end{equation}
where $\tilde{g}=\frac{2}{3\sqrt{3}} v_F \left(2\pi g^4 /p_0 \right)^{1/3}=
\frac{4\pi}{3} \omega_0^{1/3}$ is the effective interaction constant.
Note that $k_\parallel$ does not appear because it is negligible relative to
$\omega$ for the reason given below eq. (\ref{k_omega}).

After these transformations diagrams which do not contain fermion loops (except
for those loops implicit in the gauge-field propagator) become the same as in
a 1D theory with a retarded interaction given by 
(\ref{D^(1D)}) and the diagrams which contain loops are negligible.
Therefore, in the limit $N \rightarrow 0$ the theory can be solved by
bosonization methods.
Moreover, by reproducing this solution using the diagram technique we find
that one 
dimensional results depend crucially on the exact cancellations specific to
1D models and that at any $N \neq 0$ these cancellations are not exact.
These observations allow us to obtain 
some information about the behavior at $N \ll 1$.
The analysis of $N\rightarrow 0$ limit is given in Section III.
The solution at $N \rightarrow 0$ turns out to be very similar to the results
for the fermion propagator obtained by Khveschenko and Stamp
\cite{khveschenkostamp} via eikonal methods and by Kwon, Marston and Houghton 
\cite{houghton} via a two dimensional bosonization.
From our results we see that these calculations are
only valid in the strict $N \rightarrow 0$ limit, so that the claim
of these authors to have
calculated the low energy behavior exactly at $N=2$ or for the
half-filled Landau level is in disagreement with
our results.

A third theoretical approach involves scaling equations constructed by
eliminating high energy degrees of freedom.
J. Gan and E. Wong \cite{ganwong} derived an action for the gauge field alone
by integrating out the fermion degrees of freedom in Hamiltonian (\ref{H}) and
then showed that this action has an infra-red stable weak coupling fixed point
in 2 spatial dimensions.
From this they concluded that Eq. (\ref{D}) gives the correct asymptotic form of
the gauge field propagator.
Kwon et al \cite{houghton} obtained the same result via bosonisation.
Our results for finite $N$ imply that ``correct asymptotic form'' means that
the scaling $\omega \sim q^3$ is preserved, as is the behavior in the limits
$k \gg k_\omega$ and $k \ll k_\omega$, but not the precise functional form when
$k \sim k_\omega$.
An alternative scaling treatment was given by Nayak and Wilczek
\cite{nayakwilczek}, extending 
previous work of Shankar \cite{shankar} on short range interactions.
Nayak and Wilczek wrote a scaling relation for an action based directly on Eq.
(\ref{H}).
They concluded that for the $\nu=1/2$ problem in $d=2$  the fermion gauge
field interaction is marginal and in the ``spin liquid'' case it is relevant,
so that no statements can be made until the strong coupling fixed point is
found. 
However, our results imply that the strong coupling fixed point has a
straightforward interpretation: in the ``spin liquid'' case in $d$ spatial
dimensions the bare scaling $\epsilon \sim vk \sim v_F k_\perp^2/(2p_0)$ is replaced by the new
scaling $\epsilon^{d/3} \sim v_Fk_\parallel \sim v_F k_\perp^2/(2p_0)$ found 
from the leading order gauge field corrections to the fermion propagator.
In $d > 2$ we show that any additional corrections from the fermion-gauge-field
interactions are irrelevant.
In $d=2$ we show that the corrections are marginal at $x>0$ and lead to new 
power laws only in the $2p_F$ susceptibilities.
In the case of half-filled Landau level ($x=0$) these power laws are replaced
by a much weaker singularity.  
For the case of the half-filled Landau level with unscreened Coulomb 
interaction our results amount to a justification of the leading-order
approach of Halperin et. al. \cite{halperinleeread}.
The interpretation of our results in terms of scaling theory is discussed in
Section V.

Section VI is a conclusion in which the physical interpretation of our results
is discussed.

After this manuscript was completed we learned of two preprints reporting
results very similar to some of those reported here.  Kim, Furusaki, Wen
and Lee \cite{kim} calculated particle-hole bubbles at small q to 
order $1/N^2$ in the spin liquid model, finding, as we did, that the fermi
liquid form is not modified by the gauge interaction.  Polchinksi 
\cite{polchinski} performed
a scaling analysis of the large-N spin liquid model and concluded, as
do we, that the curvature of the fermi surface is important and that
Migdal-type arguments justify the results of the leading order
perturbation theory calculation.  He also obtained our result, eq. (\ref{W}),
for the renormalization of the $2p_F$ component of the four fermion 
interaction.

\section{Large N Limit}

This section will show that in the limit $N \rightarrow \infty$ the leading
contribution is given by the diagrams with the minimal number of crossings;
this will allow us to construct a perturbative series in $1/N$ and obtain
physical results in the leading orders of this expansion.
We find that to all orders in the expansion the self energy remains proportional to $\epsilon^{2/3}$, that
all particle-hole susceptibilities except those at $|{\bf Q}|=2p_F$ 
retain the usual Fermi liquid form and that 
correlators at $2p_F$ momentum transfer acquire an anomalous power law 
dependence.

\begin{figure}

\centerline{\epsfxsize=6cm \epsfbox{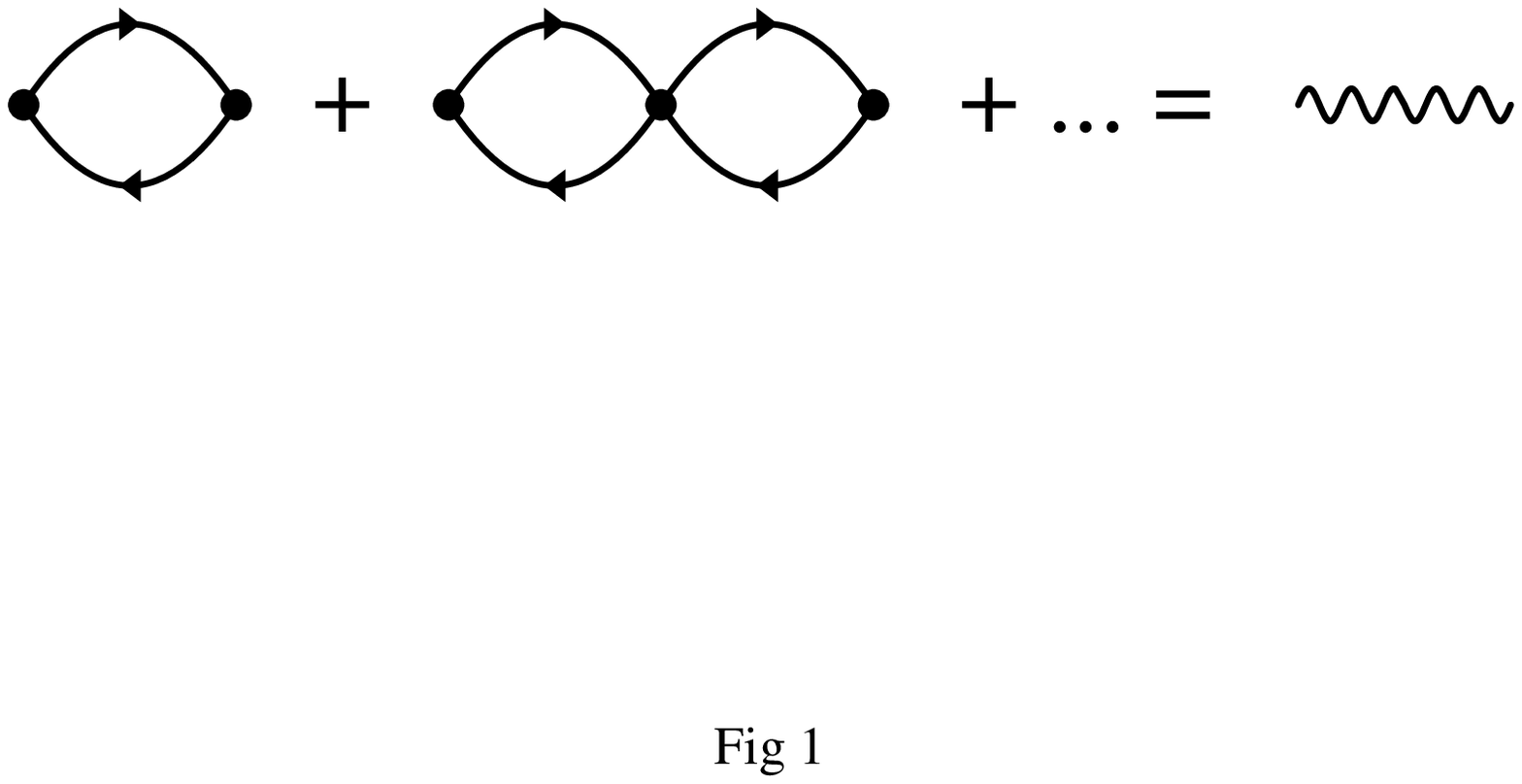}}
\caption{RPA sum of bubble diagrams leading to dressed gauge field propagator
(denoted by thick wavy line). 
The solid lines with arrows denote fermion propagators and the heavy dots
denote the bare gauge field propagator.
Whether the fermion propagators are bare or renormalized does not affect the
result of the calculation.}
\label{F1} 
\end{figure}

In order to develop a consistent large $N$ expansion for the Hamiltonian
(\ref{H})  we must 
take $N \rightarrow \infty$ limit so that the interaction parameter 
$g^2$ in (\ref{D}) remains constant.
At $N=\infty$ the only diagrams that survive are the RPA bubble graphs shown in
Fig.~\ref{F1}. These bubbles  screen the $1/k^2$ behavior of the gauge field.
Because the gauge field is transverse, it is not completely screened and the
result is Eq. (\ref{D}).

\begin{figure}
\centerline{\epsfxsize=6cm \epsfbox{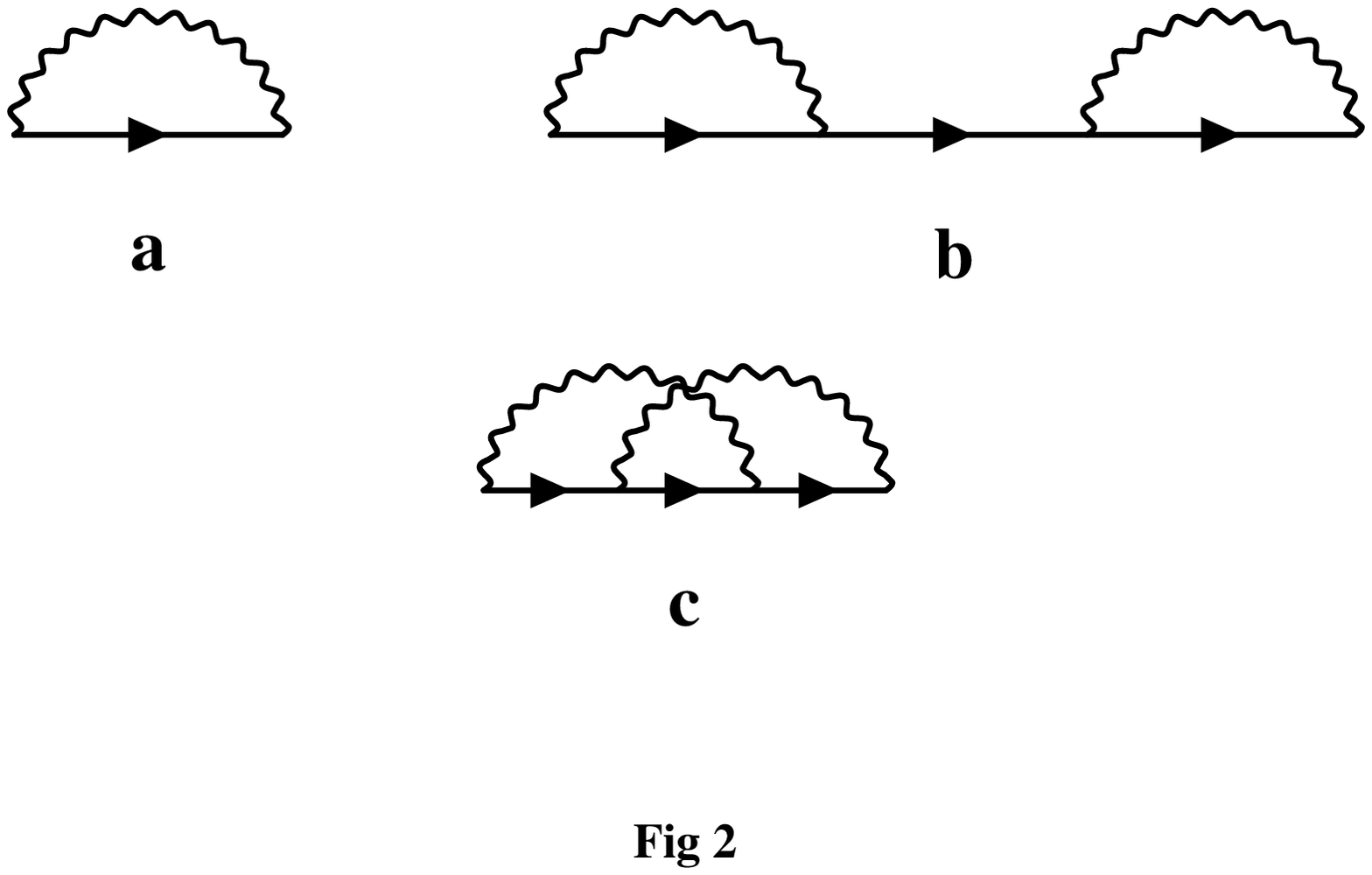}}
\caption{Fermion self energy diagrams.
The wavy line denotes the gauge field propagator (2) and
the solid line the fermion propagator.}
\label{F2} 
\end{figure}

We now consider the $1/N$ corrections to the fermion propagator.
These are shown diagrammatically in Fig.~\ref{F2}.
The self energy appearing in the leading diagram (Fig.~\ref{F2}a) was given in
Eq. (\ref{Sigma^(1)}).
One sees that at energies less than $\omega_0$ or length scales longer than
$v_F/\omega_0$, the self energy becomes larger than the inverse of the
bare Green function.
We have chosen the way the limit $N \rightarrow \infty$ is taken so that 
the scale $\omega_0$ remains constant.
Because the first correction is of the order of $1$ and not of $1/N$,
care is required in carrying the $1/N$ expansion to higher orders.

Now consider the $O(1/N^2)$ terms. 
The first of these (Fig.~\ref{F2}b) scales as 
$\left[\Sigma^{(1)}(\epsilon)\right]^2/\epsilon \sim \epsilon^{1/3}$.
Direct calculation using bare fermion propagators shows that the second term
(Fig.~\ref{F2}c) scales as $\epsilon$ (up to logarithms). 
Specifically:
\begin{equation}
\Sigma^{2}_b(\epsilon,p_\parallel)=\frac{c}{(2\pi)^2 N^2} 
(i\epsilon - v_Fp_\parallel) 
\left( \ln \frac{\epsilon^{2/3} \omega_0^{1/3}}{|\epsilon + iv_Fp_\parallel|}
\right)^2
\end{equation}
where $c \approx 3.28$ and $p_\parallel=|p|-p_F$.
This shows that in the low energy limit the self energy is more singular than
the vertex correction and should be summed first.
To calculate to higher order in the $1/N$ expansion we should therefore use the
Green function $G^{(1)}$ given by
\begin{equation}
G^{(1)}(\epsilon,{\bf p}) = \frac{1}{i\epsilon-v_Fp-\Sigma^{(1)}(\epsilon)},
\label{G^(1)}
\end{equation}
with the self enegy $\Sigma^{(1)}(\epsilon)$ is given by (\ref{Sigma^(1)}).
In fact this $G^{(1)}$ solves the self-consistent Eliashberg equation
$\Sigma=\int D G$ also, because $\Sigma^{(1)}(\epsilon)$ is momentum
independent \cite{migdal}.  Therefore,  the rainbow graphs have been summed and we need only to
consider graphs with crossed lines such as shown in Fig.~\ref{F2}c.

\begin{figure}
\centerline{\epsfxsize=6cm \epsfbox{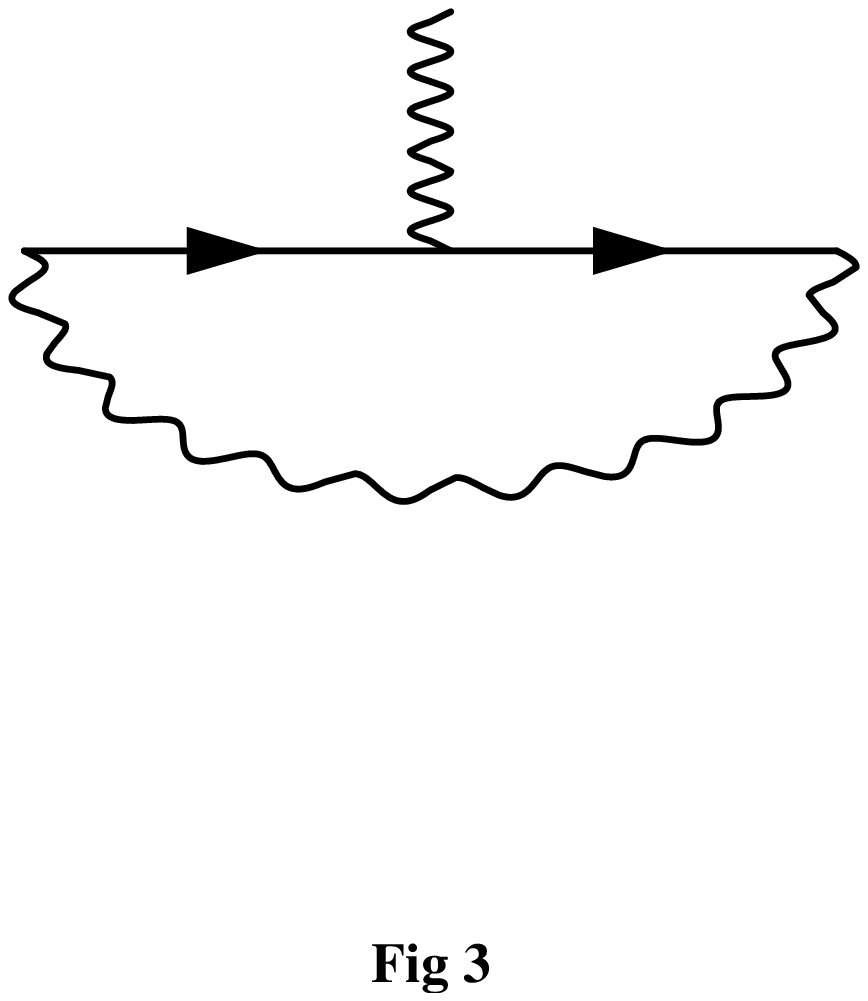}}
\caption{Correction to fermion-gauge field vertex.
The wavy line denotes the gauge field propagator (2)  and
the solid line the fermion propagator.}
\label{F3} 
\end{figure}

Returning now to the vertex corrections we reevaluate the leading vertex correction, shown in
Fig.~\ref{F3}a, using (\ref{G^(1)}) for the fermion Green functions. 
We find that this correction is at most of the order of the bare 
vertex, moreover, it is of order $(\ln [N]/N)^2$ for  external momenta of order
$k_\omega$. 
Explicitly, we find   
\begin{eqnarray}
\Gamma_{p,0}^{(1R)}(\omega,q) =&& \frac{\sqrt{3}v_F}{\pi}  
	F\left[ 2^{1/3} \sqrt{3} \pi N \!
	\frac{q_{\perp}}{k_\omega} \right]
\label{Gamma^(1R)} \\
&& \begin{array}{ll}
	F(x)=\frac{2^{1/3} 3}{4} \frac{1}{x^2} \ln^2\left(\frac{1}{x}\right) 
						\;             & x\gg1\\
	F(x)=0.9422                                            & x=0
\end{array} 
\nonumber
\end{eqnarray}

Qualitatively, the small value of the vertex correction at large $N$ can be
attributed to the argument underlying the Migdal theorem in the
electron-phonon problem \cite{migdal}, namely that the ``velocity'' of the 
boson is much
less than the ``velocity'' of the electron (by ``velocity'' in the present
case we mean $\omega/|k|$).
However, the argument is more subtle than in the electron-phonon problem 
because here we have only small angle scattering.
To understand how the argument goes, consider again the second order crossed
graph for the self energy Fig.~\ref{F2}c, using now (\ref{G^(1)}) for
the fermion Green function.

\end{multicols}
\begin{eqnarray}
\Sigma^{(2)}(\epsilon,p) &=& v_F^4 
\sum_{\omega_1,\omega_2} \int G^{(1)}(\epsilon+\omega_1,p+k_1)
    G^{(1)}(\epsilon+\omega_2,p+k_2)
G^{(1)}(\epsilon+\omega_1+\omega_2,p+k_1+k_2)
\nonumber \\
&\times& D(\omega_1,k_1) D(\omega_2,k_2) (d^2k_1d^3K_2)
\label{Sigma2}
\end{eqnarray}
In order to evaluate (\ref{Sigma2}) we integrate over the parallel components
of the momenta $k_{1\parallel}$ and $k_{2\parallel}$ obtaining:

\begin{equation}
\Sigma^{(2)}(\epsilon,p_\parallel) = v_F^2 
{\sum_{\omega_1,\omega_2}}' \int \frac{D(\omega_1,k_{\perp1})
D(\omega_2,k_{\perp2})  
(dk_{\perp1} dk_{\perp2})}{ A + \frac{v_F}{p_0} k_{\perp1} k_{\perp2}}
\label{Sigma2_1}
\end{equation}
where the prime means that the sum over frequencies is restricted to the region 
where $\omega_1 +
\omega_2 + \epsilon$ has sign opposite to $\omega_1+\epsilon$ and $\omega_2 +
\epsilon$ and
\begin{equation}
A(\omega_1,\omega_2,p_\parallel) = v_F p_\parallel + 
i \omega_0^{1/3} \left( 
|\epsilon+\omega_1 +\omega_2|^{2/3} + |\epsilon + \omega_1|^{2/3} + 
|\epsilon + \omega_2|^{2/3} \right)
\end{equation} 
\begin{multicols}{2}

Clearly, the second order contribution to the self energy is at most of the 
order of $1/N$ because it contains 
$1/N^2$ coming from two gauge propagators and $N$ from the phase volume (note
that $k_\omega \propto N^{1/2}$).
In fact, the coefficient of the $1/N$ term vanishes because the expression
under the integral in (\ref{Sigma2_1}) is odd in $k_{1\perp}$ and $k_{2\perp}$
and the leading behavior turns out to be 
\begin{equation}
\Sigma^{(2)}(\epsilon,p_\parallel=0) 
= - i c' \left( \frac{\ln N}{4\pi N} \right)^2
	\epsilon \left|\frac{\omega_0}{\epsilon}\right|^{1/3}, 
\hspace{0.25in} c' \approx 2.16
\label{Sigma^(2)_2}
\end{equation}
The reason for the powers of $1/N$
is essentially that the phase volume available for
the process when all three electron lines are on the mass shell is negligible
as in the usual Migdal arguments \cite{migdal}, although here the phase volume is small only
in $1/N$.
Note that the non-zero curvature of the Fermi surface is essential to the 
argument.
Note also that in spatial dimension $d>2$ the leading self energy is
$\epsilon^{d/3}$ so that at any N the small parameter of the ``Migdal expansion'' is
$\epsilon^{\frac{d-2}{3}}$.
This is related to the fact, to be discussed at greater length in Section V,
that the interaction is marginal in $d=2$ and irrelevant in $d>2$.
Note that although $\Sigma^{(2)}$ has the same form as $\Sigma^{(1)}$ in the
limit $\omega_0^{1/3} |\epsilon|^{2/3} \gg v_F p_\parallel$ it does not have
exactly the same functional form for
$\omega_0^{1/3} |\epsilon|^{2/3} \sim v_F p_\parallel$.

Thus, at $N \gg 1$ all diagrams can be classified by the number of crossings
and the sets of diagrams with minimal number of crossings should be summed
first, a procedure well known from localization theory
\cite{altshuleraronov}.
The result of this summation shows that such diagrams indeed give the
leading contributions to the higher order terms of the perturbation expansion 
but these contributions are not sufficiently singular at low energies and 
contain extra powers of $1/N$. 
We discuss the calculations leading to this conclusion in Appendix B.

The absence of low energy singularities in the higher orders of the
perturbation theory implies that the results obtained in the leading order are
modified only slightly by higher order terms.

The discussion so far has shown that the $1/N$ expansion is well
defined and has established the qualitative form of the fermion propagator. 
Now we verify that higher order corrections in $1/N$ do not change the
qualitative form of the gauge field propagator.
This follows from the general considerations of Gan and Wong \cite{ganwong}, 
but we believe
an explicit derivation is valuable because the validity of the approach
of Gan and Wong (which
involved integrating out gapless fermions and dealing with an action involving the gauge field only) may be questioned and because the derivation makes clear that although
the two limits ($k \gg k_\omega$ and $k \ll k_\omega$) are correctly given by
Eq. (\ref{D}), the precise form for $k \sim k_\omega$ is changed by higher
order diagrams.

We first consider the leading term ${D^{(R)}}^{-1}(\omega,q)$, which is 
obtained by evaluating the polarization bubble (Fig 1) but with renormalized 
fermion propagators.
This may be written 
\end{multicols}
\begin{equation}
{D^{(R)}}^{-1}(\omega, q) = \sum_{\epsilon, p}
\frac{1}{i \omega_0^{1/3}|\epsilon+\omega|^{2/3}-v_Fp_\parallel - 
\frac{v_F}{2p_0}(p_\perp+q)^2} \;  
\frac{1}{i \omega_0^{1/3}|\epsilon+\omega|^{2/3}-v_Fp_\parallel - 
\frac{v_F}{2p_0}p_\perp^2} 
\label{D^(R)}
\end{equation}
\begin{multicols}{2}

\begin{figure}
\centerline{\epsfxsize=6cm \epsfbox{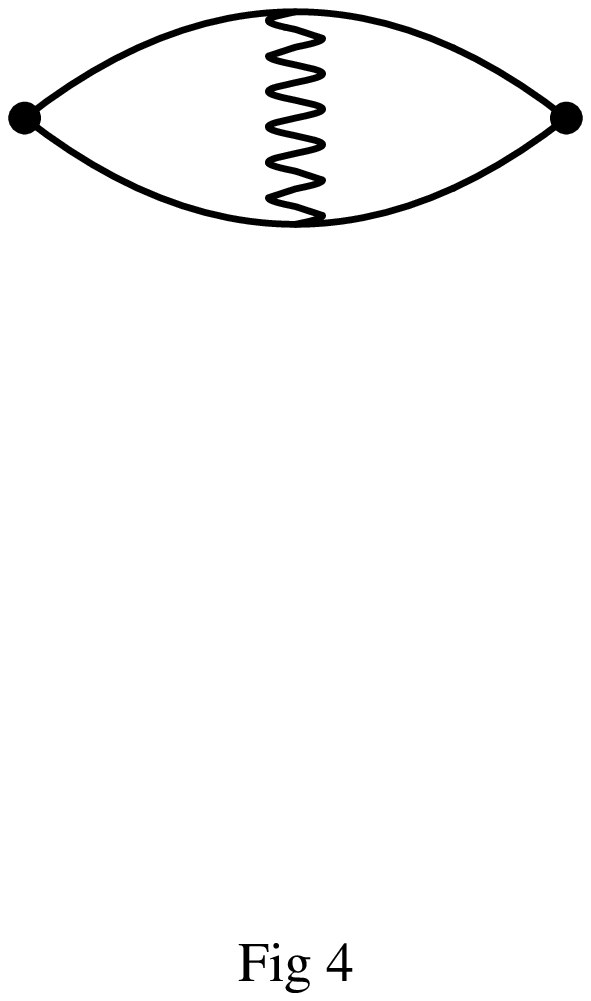}}
\caption{Correction to fermion polarizability. 
The wavy line denotes the gauge field propagator (2) and
the solid line the fermion propagator.}
\label{F4} 
\end{figure}

This may be most easily evaluated by subtracting and adding the bare bubble
obtained using bare Green functions in (\ref{D^(R)}).
In the difference term one may integrate over $p_\parallel$ first, then sum
over $\epsilon$.
The result is
\begin{equation}
D_{(R)}^{-1}(\omega, q) - D^{-1}(\omega, q) 
	\sim \frac{p_0 \omega_0^{1/3} |\omega|^{5/3}}{v_F q^2}
\label{deltaD}
\end{equation}
Thus, for $|\omega| \ll \omega_0^{-1/2} (v_F q)^{3/2}$, the full propagator is
still of the bare form (\ref{D}).
Further, we only need this propagator for $\omega \sim N^{-3/2} g^{-2}
p_0^{-1} q^3 \ll \omega_0^{-1/2} (v_F q)^{3/2}$.
Thus, the only effect of using renormalized Green functions is to reduce the
upper frequency cutoff (which enters no physically interesting result) from
$vq$ to $\omega_0^{-1/2} (v_F q)^{3/2}$.
A very similar calculation shows that $1/N$ vertex correction to the
polarization bubble shown in Fig.\ref{F4}a is of the same order as the self
energy correction to the bubble, i.e.: 
\begin{equation}
\delta D(\omega,q) \sim \frac{p_0 \omega_0^{1/3} |\omega|^{5/3}}{v_F q^2} 
\end{equation}
Thus, for $\omega$ less than the upper cutoff $\omega_0^{-1/2} (v_F q)^{3/2}$
the renormalization of the gauge field propagator is small.
In particular, for $q \sim k_\omega$ it is smaller than the bare
part by a factor of order of $q$.
However, the two loop diagram, Fig.~\ref{F4}b, leads to a correction which is
of the same order as the leading diagrams (Fig.~\ref{F1}) in the infrared
limit.
We do not discuss the details of the evaluation here (except to note that the
dominant contribution comes when the internal gauge-field momentum $q$ is 
almost parallel to the external momentum, $k$, and that the fermion loop
vanishes if the external frequency $\omega=0$ but are of the order of unity if
$k \sim k_\omega$).
This diagram therefore does not change the scaling or the asymptotic forms 
in the limits $k \gg k_\omega$ or $k \ll k_\omega$ but does change the
detailed $\omega, k$ dependence of $D(\omega,k)$.
This discussion also shows that in general the long wave susceptibilities
preserve the Fermi liquid form for small frequencies.

So far we have discussed the effects of gauge fields on the long wave properties of
fermions.
Now we turn to the effects of gauge field on the fermion vertices with large
momentum transfer.
The corrections to the vertex with large but arbitrary momentum transfer 
$|{\bf q}| \sim p_F$ are generally small because of the small phase volume 
available for virtual processes which leave both fermions with momentum 
transfer $\bf p+q+k$ and  $\bf p+k$ close to the Fermi surface.
The situation changes only for $|q|$ close $2p_F$.
In this case a virtual process with momentum transfer $\bf q$ along the 
Fermi surface leaves both fermions with
momenta $\bf p+q+k$ and $\bf p+k$ near the Fermi surface.

The leading  contribution in $1/N$ to the fermion vertex $\Gamma_{Q}$ is
logarithmically divergent at $Q=2p_F$;
we find that higher powers of $N$ contain higher powers of logarithms;
we sum these logarithms using a renormalization group method and find
power law singularities in $\Gamma_{2p_F}$.
These singularities imply that the calculation of the particle-hole
susceptibility must be reconsidered.
Finally, a singular susceptibility near $2p_F$ may be further modified by
the short range four fermion interaction; therefore we must consider
also the renormalization of this interaction by the gauge fields.

We begin with the diagrams for $\Gamma_Q$ shown in Fig.~\ref{F5}.
The diagrams shown there diverge logarithmically if all external momenta are on
the  Fermi surface, external energies are zero and the momentum transfer is 
exactly $Q=2p_F$.
Since the energy only enters the Green function via
$\omega_0^{1/3}|\epsilon|^{2/3}$, the momentum component across the Fermi 
surface via $v_Fk_{\parallel}$ and the momentum along the Fermi surface
via $k_{\perp}^2$, the divergence is cut off by the largest among 
\begin{equation}
\epsilon, \; \;
\epsilon_\parallel=\frac{(v_Fk_{\parallel})^{3/2}}{\omega_0^{1/2}}, \; \;
\epsilon_\perp=\frac{k_{\perp}^{3}/m^{3/2}}{\omega_0^{1/2}}.
\label{gaugescaling}
\end{equation}
If, say, the largest is the external frequency we evaluate the diagrams in
Fig.~\ref{F5} and get 
\begin{equation}
\delta \Gamma_{2p_F}(\omega) = 
	\left( \frac{1}{2N} + \frac{1}{2 \pi^2N^2} \ln^3(N) \right) 
 \ln\left(\frac{1}{\omega}\right) \Gamma^0_{2p_F}
\label{delta_Gamma} 
\end{equation}
where $\Gamma^0_{p_F}$ is the bare vertex at small scales or large frequencies.
The logarithmic nature of the corrections to the effective interaction
allows us to sum higher orders of the perturbation theory by constructing the
renormalization group equation:
\begin{equation}
\frac{d \Gamma_{2p_F}}{d \ln(1/\omega)} = \left( \frac{1}{2N} +\frac{1}{2 \pi^2N^2} 
\ln^3(N)
	\right)  \Gamma{2p_F}
\label{RG_Gamma}
\end{equation}
>From (\ref{RG_Gamma}) we see that the vertex grows at large scales as
\begin{equation}
\Gamma^{R}_{2p_F} \sim 
\left( \frac{\epsilon_F}{\omega} \right)^{\sigma} \Gamma^0_{2p_F} 
\label{Gamma^R_2p_F}
\end{equation}
\begin{equation}
	\sigma=\frac{1}{2 N} + \frac{1}{2 \pi^2 N^2}\ln^3(N) + 
	O\left( \frac{1}{N^2} \right)
\label{sigma_1}
\end{equation}
Here we used energy $\omega$ for the infra-red cut off assuming that it sets the
largest scale among $\omega$, $\epsilon$, $\epsilon_\parallel$, 
$\epsilon_\perp$.
The result (\ref{Gamma^R_2p_F}) is derived using a large $N$ expansion. 
It is also of interest to evaluate these diagrams at $N=2$.
The leading order diagram gives $\sigma=0.25$; the sum of the diagrams shown 
in Fig.~\ref{F5}b and d gives $\sigma=0.35$.

The power law growth of the vertex at $2p_F$ distinguishes fermions with a gauge
interaction from an ordinary Fermi liquid with short range repulsion and
leads to anomalous behaviour of the spin correlators at $Q=2p_F$.
In the absence of a short range interaction effective at $2p_F$ (i.e. if
the interaction W in eq. \ref{H} vanishes) the spin
correlator is given by the polarization diagrams shown in Fig.~\ref{F6}.
The leading contributions in powers of $(\frac{1}{N}\ln\omega)$ come from the 
diagrams in which the vertical lines 
of the gauge field do not cross. 
In these diagrams the leading contribution originates from 
the frequency range (and corresponding
momentum range, which we have not explicitly written)
\[
\epsilon_F > \omega_n > \cdots  \omega_1 > \omega < \cdots 
< \omega_{-n} < \epsilon_F
\]
where $\omega$ is external frequency.
Therefore, the sum of all diagrams is given by the diagram shown in
Fig.~\ref{F6} with renormalized vertices (\ref{Gamma^R_2p_F}):
\end{multicols}
\begin{equation}
\Pi(\omega,q) = \int G(\epsilon+\omega/2,p+q/2) G(\epsilon-\omega/2,p-q/2)
[\Gamma^{(R)}_{\epsilon,p}(\omega,q)]^2 (dp d\epsilon)
\label{Pi}
\end{equation}
To evaluate the integral in (\ref{Pi}) we note that the main contribution to
it comes from the range of momenta and energies related by $\epsilon \sim
\epsilon_\parallel \sim \epsilon_\perp \sim \omega$ (\ref{gaugescaling}). 
Estimating the result by power counting we find that if $\sigma < 1/3$ (as
occurs for large $N$) the integral (\ref{Pi}) converges, but if $\sigma > 1/3$
it diverges at $\omega=0$, $q=2p_F$. 
We evaluate the integral in these cases separately and find:
\begin{eqnarray}
\Pi(\omega,q) = \Pi_0 -  \sqrt{\frac{p_0}{\omega_0 v_F^3}} 
	\left[ 
	c_\omega \left(\frac{\omega}{\omega_0}\right)^{\frac{2}{3}-2\sigma} +
		c_q \left(|q-2p_F| l_0 \right)^{1-3\sigma} 
	\right] 
	&\hspace{0.5in}  & \sigma < 1/3 
\label{Pi_2D} \\
\Pi(\omega,q) = \sqrt{\frac{p_0}{\omega_0 v_F^3}} 
	\left[
	c_\omega \left(\frac{\omega}{\omega_0}\right)^{2\sigma -\frac{2}{3}} +
	c_q \left(|q-2p_F| l_0 \right)^{3\sigma-1}
	\right]^{-1}
	&\hspace{0.5in}& \sigma > 1/3
\label{Pi_1D}
\end{eqnarray} 
where the coefficients $c_q$ and $c_\omega$ are of the order of unity for a 
curved Fermi surface. Below we shall assume that $c_q=c_\omega=1$.
Since these coefficients depend strongly on the curvature of the
Fermi  surface, the case of a flat Fermi surface should be considered 
separately.  
We do not discuss it further here.
\begin{multicols}{2}

The spin polarization bubble (\ref{Pi_1D}) is equal to the spin susceptibility
if the effects of the short range interaction on the spin correlators at
$2p_F$ can be neglected.
We justify this by showing that a sufficiently weak bare interaction 
is renormalized to zero by the gauge field.
Renormalization of the effective interaction by the gauge field occurs in the
two competing channels shown in Fig.~\ref{F7}.
In both channels the corrections diverge logarithmically if all external 
momenta are on the Fermi
surface, external energies are zero and the momentum transfer is exactly 
$Q=2p_F$.
This divergence is again cut off by the largest among $\epsilon$,
$\epsilon_\parallel$ and $\epsilon_\perp$.
If, say, the largest is the external frequency we evaluate the diagrams in
Fig.~\ref{F7} and get
\begin{equation}
\delta W(\omega) = - 2 (\frac{2}{3} - \frac{1}{2N}) 
	\ln\left(\frac{1}{\omega}\right) W_0
\label{deltaW}
\end{equation}
where $W_0$ is the bare interaction at small scales or large frequencies.
Using the renormalization group to sum higher orders we conclude that
interaction at $2p_F$ decays rapidly at large scales:
\begin{equation}
W \sim \left( \frac{\omega}{\omega_0} \right)^{\frac{4}{3} - \frac{1}{N}}
	W_0
\label{W}
\end{equation}

Certainly, Eq.~(\ref{W}) holds only for sufficiently small bare
interaction $W_0$.
The decay of the interaction implies that the spin susceptibility
$\chi(\omega,q)=\Pi(\omega,q)$.  
For larger bare interaction $W_0 \geq W_c$ we expect a transition into an 
ordered  state to occur.
We will give the theory of this transition in a separate paper.
Here we note that at $W_c$ the interaction does not scale and that the
basic ingredients of the theory of the transition are polarizability of the 
fermion system (\ref{Pi_2D},\ref{Pi_1D}) and the four spin fluctuation
interaction shown in Fig.~\ref{F8}.
The renormalization of the four spin interaction may be treated in the same
way as that of $\Pi$.
We find that at large $N$ the leading diagrams are those shown in
Fig.~\ref{F8}; these lead to a divergent $U$ with the divergence cut off by
the largest among $|\omega/\epsilon_F|^{2/3}$, $q_\parallel/p_F$ and
$q_\perp/p_F$:

\begin{equation}
U(\omega_1,\omega_2,q_1,q_2) = \frac{\epsilon_F p_F^2}
{\left( \frac{\omega}{\epsilon_F} \right)^{4\sigma + \frac{2}{3}} \!+\!
\left( \frac{|q_{\perp}|}{p_F} \right)^{12 \sigma +2} \!+\!
\left( \frac{q_{\parallel}}{p_F} \right)^{6 \sigma +1} }
\label{U}
\end{equation}
Here we denote $q_{\parallel}=max(||q_1|-2p_F||, ||q_2|-2p_F||)$, 
$\omega=max(|\omega_1|, |\omega_2|)$ and 
$q_{\perp} = ({\bf q_1-q_2}) \hat{{\bf q}}$ where $\hat{{\bf q}}$ is the unit
vector in the direction of $\bf q_1+q_2$.

To summarize: in the limit of large $N$ the physical properties of the spin
liquid resemble conventional Fermi liquid with the following important 
differences: 
(i) the scaling relation between energy and momentum is changed to
(\ref{gaugescaling}),
(ii) spin correlators acquire anomalous power behavior (\ref{Pi_2D}) at
$2p_F$,
(iii) interaction vertices with external field at momentum $2p_F$ are strongly
enhanced, but 
(iv) the short ranged interaction between quasiparticles is suppressed.

\section{Small N limit}

In this Section we shall show that in the limit $N\rightarrow 0$ the motion of
fermions becomes essentially one dimensional and apply methods borrowed from 1D
theories to obtain physical results which turn out to be qualitatively similar
to the results obtained in the limit $N\rightarrow \infty$.
For the fermion propagator the $N\rightarrow 0$ limit is not singular and the
1D theory gives qualitatively the same result as the $N\gg 1$ 2D calculation.
We find that for the vertex function the  $N\rightarrow 0$ limit is singular 
because it predicts an exponentially divergent vertex function rather than the 
power law derived in the limit $N \rightarrow \infty$.
We shall show that the power law behavior remains valid for all finite 
$N$, but the power tends to infinity as $N\rightarrow 0$.

The $N \rightarrow 0$ limit is defined via equation (\ref{D}) for
$D(\omega,k)$.
In this formula we take $N$ to zero with $g^2$ constant.
To see why this $N \rightarrow 0$ limit is essentially one dimensional
consider again the second order (in the gauge field propagator) contribution 
to the fermion self energy shown in Fig.~\ref{F2}c.
Let us perform the integration over $k_\perp$ first.
In the limit $N\rightarrow 0$ the $k_\perp$ dependence of the diagram is
controlled by the gauge propagators, which implies that the main contribution
is at $k_\perp \sim N^{1/2} \epsilon^{1/3}$; this means that $k_\perp^2 \sim
 N \epsilon^{2/3}$ is negligible compared to the self energy of the fermion
($\Sigma \sim \epsilon^{2/3}$).
Therefore, at $N \rightarrow 0$ one may neglect the $k_\perp$ dependence of
all Green function lines.
In addition we may neglect all diagrams except those in which all gauge field
lines connect two electrons moving either nearly parallel or nearly 
antiparallel.
The reason is that if the gauge field couples two fermions moving
in arbitrary directions all components of the transferred momentum are limited
by fermion Green functions and become small:
$v_F |k| \sim \omega_0^{1/3} |\epsilon|^{2/3}$.
This decreases the phase space volume to $k_\parallel k_\perp \sim 
\epsilon^{4/3}$ (instead of $\epsilon$) and decreases the 
gauge field propagator to $\sim \epsilon^{-1/3}$.
As a result these processes are small (of relative order $\epsilon^{2/3}$) 
and irrelevant in the infrared limit.

Very similar arguments apply to diagrams containing internal fermion loops
other than those contributing to the gauge field propagator. 
Here the reason is that the gauge field propagator is small in $N^{1/2}$.
In the calculation of the fermion propagator this smallness was compensated by
the infrared divergence of the integral over momentum component perpendicular
to the Fermi velocity.
This divergence was cut off by $k_\perp \sim N^{1/2} \omega^{1/3}$ cancelling
the factor of $N^{1/2}$.
In the diagrams containg fermion loops, momenta in different loops are not
exactly parallel ($k_\perp \sim \epsilon^{1/3}$) and integrals over 
$k_\perp$ are cut off at $k_\perp \sim \epsilon^{1/3}$; this infra-red cut off
does not contain compensating factors of $N$.

In the remaining diagrams one may integrate the gauge field lines over the
components of momentum perpendicular to the Fermi velocity and obtain a one 
dimensional theory of electrons (with propagator depending on frequency and 
one component of momentum) coupled to the momentum independent but retarded 
interaction $D^{1D}(\omega)$ defined in (\ref{D^(1D)}).
As we shall show below, the resulting theory can be solved by bosonization
methods. 
We shall then use Ward identites to obtain information about the behavior at small
but non-zero $N$.

Therefore, in the limit $N\rightarrow 0$  the sum of all diagrams can be 
found from a mapping to a 1D theory \cite{ioffelidsky} with the action
\end{multicols}
\begin{equation}
S =\int \left( 
	\bar{\Psi}_{-\omega,-k}^{R,a}(i\omega-v_F k )\Psi_{\omega,k}^{R,a} +
	\bar{\Psi}_{-\omega,-k}^{L,a}(i\omega+v_F k )\Psi_{\omega,k}^{L,a} +
	\frac{v_F g |\rho_{\omega,k}^R-\rho_{\omega,k}^L|^2 }{|\omega|^{1/3}}
       \right) (dk d\omega)
\label{SL}
\end{equation}
\begin{multicols}{2}
Here $\rho_{t,x}=\bar{\Psi}_{t,x}^a\Psi_{t,x}^a$ is the density operator and
$a$ is a replica index which runs over $K$ values. 
We take the limit $K \rightarrow 0$ to exclude fermion loops, which we 
have argued to be negligible. 
In a conventional one dimensional theory with short range interactions loops
would be present and would affect the values of the exponents.
Our $N \rightarrow 0$ limit is defined so that loops are negligible in the
$d=2$ gauge problem.
If loops are {\em not} negligible in the $d=2$ gauge problem, their effect is
{\em not} correctly given by the 1D theory. 

To compute the fermion propagator we may restrict our attention to the right
moving  particles.
The theory is then the Tomonaga model with a retarded interaction,
and has been solved by bosonization \cite{ioffelidsky} yielding 
\[
G({\bf p},\epsilon)=\int G(x,t) e^{i\epsilon t - i(|{\bf p}| - p_F)x} dt\, dx,
\]
\[
G(x,t)=\frac{i}{2\pi (x-i v_Ft)} \exp \left(\frac{-\Gamma(2/3) l_0^{-1/3}|x|}
        {(|x|-isgn(x)v_Ft)^{2/3}} \right), 
\]
In the limit of low energy and momenta close to the Fermi surface 
$G({\bf p},\epsilon)$ acquires a simpler scaling form
\begin{equation}
G^{(1D)}(\epsilon,p)=\frac{-1}{v_F(p-p_F)}g
\left(\frac{\Gamma(2/3)l_0^{-1/3}\epsilon^{2/3}}{v_F^{2/3}(p-p_F)} \right)
\label{G^(1D)}
\end{equation}
\[
g(u)=\frac{3}{2} \exp[(-1)^{3/4}u^{3/2}] - \frac{3\sqrt{3}i}{4\pi}
	\int_0^\infty \frac{\exp[-(uy)^{3/2}] dy}{y^2 + iy -1}
\]
Although the Green functions (\ref{G^(1D)}) and (\ref{G^(1)}) have completely 
different analytical structures their qualitative properties are similar: 
both are equal to $1/v_F|p_F-p|$ in the limit 
$\omega_0^{1/3} |\epsilon|^{2/3} \ll v|p-p_F|$ and both behave as 
$1/(\omega_0^{1/3} |\epsilon|^{2/3})$ in the opposite limit 
$\omega_0^{1/3} |\epsilon|^{2/3} \gg v|p-p_F|$.
We therefore expect a smooth crossover from formula (\ref{G^(1D)}) 
to (\ref{G^(1)}) as $N\rightarrow 0$. Both describe overdamped fermions 
with a characteristic energy that scales as $(p-p_F)^{3/2}$.
Thus, the limit $N\rightarrow 0$ is not singular for the fermion Green
function.
Khveschenko and Stamp \cite{khveschenkostamp} obtained via eikonal methods a 
form very similar to (\ref{G^(1D)}).
They claimed their result was asymptotically exact for all $N$.
Our derivation, on the other hand, suggests that the precise form depends on
two special ``one-dimensional'' features: the neglect of internal loops and
the neglect of the perpendicular momentum in fermion propagators.
Both these features are present in the $N\rightarrow 0$ limit and in eikonal
approximation of Khveschenko and Stamp, but are not present at arbitrary $N$.
Of these two approximations the most crucial is the neglect of the
perpendicular momentum.
If the $p_\perp$ dependence of the bare Green functions is retained the
dressed Green function will not have the exponential form (\ref{G^(1D)}).We do not give the algebra here but below we apply similar arguments to the $2p_F$ vertex.
If the $p_\perp$ dependence is neglected but loops are taken into account,
the one dimensional formalism will lead to Eq. (\ref{G^(1D)}) but with a
renormalized argument. 
For this reason we do not believe that the exponential form is generic,
although the correspondence between the $N \rightarrow 0$ and $N \rightarrow
\infty$ limits lead us to believe that the scaling $\epsilon^{2/3} \propto
p_\parallel$ is.
Kwon et al \cite{houghton} also obtained a result very similar to 
(\ref{G^(1D)}) from a two
dimensional bosonisation method in the problem of half-filled Landau
level.
Again, we do not believe the result is correct for any $N>0$.

We now consider the renormalization of the $2p_F$ vertex. 
In the strictly 1D limit $N\rightarrow 0$ all diagrams leading to this 
renormalization coincide with the diagrams of 1D Luttinger model (\ref{SL}) 
which has both right and left movers.
The Luttinger model can be solved by bosonization. 
One finds \cite{ioffelidsky} that the renormalized  vertex $\gamma^R_{2p_F}$ grows exponentially: 
\begin{equation}
\Gamma^{R}_{2p_F}(\omega) \sim 
	\exp\left(\frac{3g}{2\pi|\omega|^{1/3}}\right)
\label{Gamma^R_2pF_1D}  
\end{equation}
In order to understand the reason for such rapid growth it is convenient to 
consider the calculation diagrammatically.
In order to obtain the renormalization of the $2p_F$ vertex in a conventional 
Luttinger liquid with a short range interaction $v_{SR}$ one first notices that
the first correction to the $2p_F$ vertex is logarithmic; then it can be
proved that renormalization of $v_{SR}$ cancels with the fermion self energy
\cite{dzyaloshinskylarkin}, so that the leading contribution to the $2p_F$ 
vertex comes from the ladder sum. 
In each block of this ladder one can use bare 
vertices and Green functions; finally, the ladder sum exponentiates leading 
to a power law dependence with exponent determined by $v_{SR}$.
In the present problem the singular interaction means that the first 
correction, $\delta \Gamma_{2p_F}$, to the bare $2p_F$ vertex $\Gamma^0_{2p_F}$ is a power law,
\begin{equation}
\delta \Gamma_{2p_F} = \Gamma^0_{2p_F} \frac{3g}{2\pi |\omega|^{1/3}}
\end{equation}
but renormalization of the interaction $\frac{3g}{2\pi |\omega|^{1/3}}$ still cancels with the 
fermion self energy and the series exponentiates leading to the exponential
dependence given in (\ref{Gamma^R_2pF_1D}).
The cancellation of the interaction renormalization with the fermion self energy is 
guaranteed by the Ward identity of the 1D theory.  This relates the exact
density vertex $\Gamma^{(1D)}_{\epsilon,p} (\omega,q)$ to the exact
Green function G, and reads
\cite{dzyaloshinskylarkin}:
\end{multicols}
\begin{equation}
\Gamma^{(1D)}_{\epsilon,p} (\omega,q) =  
\frac{G^{-1}(\epsilon+\omega/2,p+q/2)-G^{-1}(\epsilon-\omega/2,p-q/2)}
{i \omega-vq}
\label{Gamma^(1D)}
\end{equation} 
This identity implies that the singular part of the product of the full Green function
and the renormalized vertex is equal to the singular part of the product of
the bare Green function and the bare vertex.

This cancellation no longer holds in two dimensions.
Instead in 2D the Ward identity is 
\begin{equation}
i \omega \Gamma^{0}_{\epsilon,p} (\omega,q) - 
q_\parallel \Gamma^{\parallel}_{\epsilon,p} (\omega,q) - 
q_\perp \Gamma^{\perp}_{\epsilon,p} (\omega,q) = 
	G^{-1}(\epsilon+\omega/2,p+q/2) - G^{-1}(\epsilon-\omega/2,p-q/2)
\label{Ward}
\end{equation} 
Here we have distinguished the density vertex $\Gamma^0$ from the two
components of the current vertex, and we have written the two components
of the current vertex in coordinates parallel and perpendicular to
${\bf v_F}({\bf p})$.  The gauge field couples to fermions via the current
vertex.

In the one dimensional Tomonaga model $\Gamma^{\perp}$ is absent and 
$\Gamma^{\parallel}=v_F\Gamma^0=v_F\Gamma^{1D}$ because the current is proportional to the
density for fermions moving in one direction.
In a general two dimensional theory $\Gamma^0$, $\Gamma^{\parallel}$ and
$\Gamma^{\perp}$ are not simply related; however, in the present model which
has only small angle scattering the identity 
$\Gamma^{\parallel}=v_F\Gamma^0$ is still valid
up to terms of the order of $\epsilon^{2/3}$ or $q_\perp^2$.
Further, we show in Appendix A that at sufficiently small $q_\perp$,
$\Gamma^{\parallel}$ and $\Gamma^{\perp}$ are related via: 
\begin{eqnarray}
\Gamma^{\perp}(\omega,q)&=& B(\omega,q) sgn(q_\perp) \Gamma^\parallel(\omega,q)
\label{Gamma_perp}\\
B(\omega,q) &=& \frac{v_F q_\parallel}{2|q_\perp|}  
	\left( \sqrt{1-\frac{2\alpha|\omega| |q_\perp|}
	{(v_Fq_\parallel)^2} +i0} - 1 \right)
\label{B}
\end{eqnarray}
where $\alpha=\frac{N^{1/2} v^2 g^2}{2\pi p_0}$.  The range of q over which this result applies is given in Appendix A.
 
Using (\ref{Gamma_perp}) and (\ref{B}) in Ward identity (\ref{Ward}) we find 
\begin{equation}
\Gamma_{\epsilon,p}^{(0)}(\omega,q) = 
	\frac{G^{-1}(\epsilon+\omega/2,p+q/2)-G^{-1}(\epsilon-\omega/2,p-q/2)}
	{i \omega-v_Fq_\parallel - v_F |q_\perp| B(q_\parallel,\omega)}
\label{Gamma^(0)}
\end{equation}
The vertex of the two dimensional theory differs from the one dimensional
vertex (\ref{Gamma^(1D)}) by the term proportional to $q_\perp$ in the
denominator of (\ref{Gamma^(0)}).
Although this term is small in the limit $N\rightarrow 0$, it is important
because it smears the singularity which appears at $\omega=v_Fq_\parallel$ in
the 1D theory.
>From (\ref{Gamma^(0)}) we can calculate the renormalization of the $2p_F$
vertex as was done for the strictly 1D theory.
Consider the diagram shown in Fig.~\ref{F5}a, put the external Green functions
on the mass shell and use (\ref{Gamma^(0)}). 
The result is
\begin{equation}
\delta \Gamma_{2p_F} = \int dk_\parallel dk_\perp d\omega 
	\frac{1}{|i\omega - vk_\parallel - |k_\perp| B(\omega,k_\parallel)|^2}
	\;
	\frac{1}{\frac{Np_0|\omega|}{2\pi|k_\perp|} + 
	\frac{1}{N^{1/2}g^2} |k_\perp|^2}
\label{dummy}
\end{equation}
\begin{multicols}{2}
The $k$ integral in (\ref{dummy}) is dominated by $k\sim k_\omega \propto
|\omega|^{1/3}$; for $k_\perp$ in this range we estimate
$B(\omega,k_\parallel) \sim \frac{\omega}{v_Fk}$ which implies that the
$k_\parallel$ integral is dominated by the region $k_\parallel \sim 
\sqrt{\frac{\omega k_\perp}{v_F}} \ll k_\omega$ while the main contribution to
the $k_\perp$ integral comes from the region $k_\perp \sim k_\omega$.
Combining these estimates with Eq. (\ref{k_omega}) gives
\begin{eqnarray}
\delta \Gamma_{2p_F} &\sim& \frac{1}{N} \int \frac{d\omega dk_\perp}
	{(\omega k_\perp)^{1/2}} \frac{k_\perp}{\omega
	[1+(k_\perp/k_\omega)^3]} 
\nonumber \\
 &\sim& \frac{1}{N} \int \frac{d\omega}{|\omega|} \left( \frac{k_\omega}
	{|\omega|^{1/3}}\right)^{3/2} \sim \frac{1}{\sqrt{N}} \ln \Omega
\end{eqnarray}
For a more precise calculation, including the coefficient of $1/\sqrt{N}$, see
appendix C.
These corrections exponentiate as before leading to a power law form for 
$\Gamma_{2p_F}$ with an exponent $\sigma$ which diverges as $N\rightarrow 0$.
Explicitly, we find:
\begin{equation}
\sigma=\frac{16 \sqrt{2}}{9\pi \sqrt{N}} + O(1)
\label{sigma_2}
\end{equation}
It is interesting to numerically evaluate the exponent at $N=2$. 
We find $\sigma=0.56$.

>From the result for $\Gamma_{2p_F}$ we may obtain as before an expression for
the polarization bubble if the short range $2p_F$ interaction $W$ can be
neglected. 
The calculation of $W$ is similar to that leading to Eq. (\ref{W}) in the
previous section.
One obtains a scaling equation
\begin{equation}
\delta W(\omega) = \beta(N) \ln\left(\frac{1}{\omega}\right) W + O(W^2)
\end{equation}
In a strictly 1D theory, $\beta=0$ and the leading term in the scaling
equation is proportional to $W^2$.
In our case we find for small $N$
\begin{equation}
\beta(N)=-|c| \sqrt{N}
\end{equation}
Because the $\beta$ function is negative both at small and at large $N$ we
believe it is negative at any $N$.
Therefore we may again apply the calculation which lead us to Eq. (\ref{Pi_2D})
for polarization bubble, however, since $\sigma$ diverges as $N\rightarrow 0$ ,
the result is Eq. (\ref{Pi_1D}).

\section{Half-filled Landau level}

In this section we treat the singular interaction argued \cite{halperinleeread}
to be relevant to the problem of the half filled Landau level.
The physical problem leads to two new features: a Chern-Simons term coming from
a singular gauge transformation which eliminates the explicit dependence on
magnetic field and a long range Coulomb interaction (absent in the spin liquid
case because spinons have no charge).
In previous treatments \cite{halperinleeread,kalmeyerzhang,nayakwilczek} the 
Coulomb interaction was taken to be long ranged.
We note that in many experimental situations the device containing the half
filled Landau level may also contain a metallic gate which screens Coulomb
interaction on length greater than a screening length $\kappa^{-1}$. 
The resulting gauge field propagator which includes the RPA self energy of fermion
loops and takes into account the dielectric constant $\hat{\epsilon}$ of the
host semiconductor is
\begin{equation}
\tilde{D}(\omega,k) = \frac{1}{\frac{p_0 |\omega|}{2\pi |k|} + 
	\frac{u k^2}{k+\kappa}}
\label{tildeD}
\end{equation}
Here $u=\frac{e^2}{8\pi\hat{\epsilon}}$ and the appearance of the
$1/(8\pi)$ instead of the conventional $2\pi$ may be traced to a $1/(4\pi)$ in the coefficient of the Chern-Simons 
term \cite{halperinleeread}. 

In this section we treat the case $\kappa=0$; we expect the results to apply if
the momenta of interest $k_\omega'$ are greater than $\kappa$. In the other
limit, one should use the results of the previous section
interpolated to $N=1$.
The momenta $k_\omega'$ are those for which two terms in denominator of
(\ref{tildeD}) are comparable. At temperature $T$, typical frequencies are 
$\omega=2\pi T$ and, if $\kappa=0$, we find that typical momenta $k_T' \sim
(8 \pi p_0 k_B T \hat{\epsilon}/e^2)^{1/2}$.
Using a typical Fermi momentum for $Ga-Al-As$ system $p_0=(4\pi n)^{1/2} 
\approx 8\times 10^5 \; cm^{-1}$ and a typical $\hat{\epsilon}=13$ we find 
that the unscreened results apply if 
\begin{equation}
\kappa\;[cm^{-1}] < 4 \times 10^5 T^{1/2} \;[K]
\label{kappa}
\end{equation}
Thus if at $T=0.1\;K$ the screening layer is further than $1000\;\AA$ from the
2d electron gas, the unscreened results apply.
If it is much closer, then one should use the results of the previous section
interpolated to $N=1$.

We turn now to computations using $\tilde{D}$ (\ref{tildeD}) with $\kappa=0$.
The leading order self energy (Fig.~\ref{F2}) is
\begin{equation}
\Sigma^{(1)}(\epsilon)= -i \frac{2 \hat{\epsilon} v_F}{\pi e^2}
\ln\left(\frac{\epsilon_F}{|\epsilon|}\right) \epsilon + \ldots
\label{Sigma^(1)_hf}
\end{equation}
Here the ellipsis indicates terms which are less singular as $\epsilon
\rightarrow 0$.
Arguments identical to those of section II show that $\Sigma^{(1)}$ also solves
the leading order Eliashberg equation, so it sums correctly all rainbow graphs.

We now argue that higher order crossed diagrams give less singular
contributions to $\Sigma(\epsilon,p)$, so that the leading dependence is given
exactly by (\ref{Sigma^(1)_hf}).
Consider the leading crossed diagram, Fig.~\ref{F2}c, with the fermion 
propagators dressed by the self energy (\ref{Sigma^(1)_hf}).
After integration over parallel momenta and symmetrization in $q_{\perp 1}$,
$q_{\perp 2}$ one finds
\end{multicols} 
\begin{equation}
\Sigma^{(2)}(\epsilon) = v_F^2 
{\sum_{\omega_1,\omega_2}}' \int 
\frac{(dk_1)}{\frac{p_0 |\omega_1|}{2\pi |k_1|} + u|k_1|} \;
\frac{(dk_2)}{\frac{p_0 |\omega_2|}{2\pi |k_2|} + u|k_2|} \;
\frac{A}{ A^2 + \frac{v_F}{p_0} k_1 k_2}
\label{Sigma^(2)_hf}
\end{equation}
with 
\[
A(\omega_1,\omega_2,p_\parallel) = v_F p_\parallel  
+\Sigma^{(1)}(\epsilon+\omega_1+\omega_2)
+\Sigma^{(1)}(\epsilon+\omega_1) + \Sigma^{(1)}(\epsilon+\omega_2) 
\]
\begin{multicols}{2}
The prime on $\sum_{\omega_1,\omega_2}$ denotes the contraint that sum over 
frequencies is restricted to the region where $\omega_1 + \omega_2 + \epsilon$
has sign opposite to $\omega_1+\epsilon$ and $\omega_2 + \epsilon$.
This constraint implies that $\omega_1$ and $\omega_2$ cannot vanish
simultaneously, so no infra-red singularities arise from the frequency
integrals.
To extract the infra-red behavior of (\ref{Sigma^(2)_hf}) we may replace $A$ by
its typical value $\epsilon \ln(\epsilon_F/\epsilon)$ and $\omega_{1,2}$ by
their typical values $\epsilon$.
The sum over frequency gives a factor of $\epsilon^2$.
The main contribution to the integrals over $k_1$, $k_2$ is a logarithmic
divergence coming from the region $\epsilon < q^2 < \epsilon \ln \epsilon$; the
final result is
\begin{equation}
\Sigma^{(2)}(\epsilon) = \frac{\hat{\epsilon} v_F}{e^2}
	\epsilon \frac{\ln^2\ln(\epsilon_F/\epsilon)}
	{\ln \epsilon_F /\epsilon }
\label{Sigma^(2)_3}
\end{equation}
This is smaller than the leading term by the factor 
\[
\left(\frac{\ln\ln(\epsilon_F/\epsilon)}{\ln(\epsilon_F/\epsilon)}\right)^2
\]
Similar considerations apply to higher order crossed graphs.

Our result, that the leading behaviour at small frequencies is given exactly by
the first order diagram, is reminiscent of the Migdal theorem \cite{migdal}, which states 
that the leading low-frequency behavior of the electron self-energy in the
electron-phonon problem is given exactly by the leading order diagram.
The physical fact underlying Migdal theorem is that the momentum transferred in
an electron-phonon process is large (of the order of $p_F$) while the energy is
small (of the order of Debye frequency and much less than $v_Fk_F$).
A very similar argument applies here.
In the calculations leading to Eq. (\ref{Sigma^(1)_hf}) the energy transferred
by the gauge field is small, while the integral over momenta is logarithmic and
only cut off at the scale $p_F$.
In the spin liquid case discussed in the previous Sections all momentum
integrals were confined to the region of small momenta.
The problem  simplified only in the large $N$ limit where the range of the momentum integration became large in $N$.
Thus, the problem of half-filled Landau level is analogous to the large $N$
limit of the spin liquid case.
Kwon et al \cite{houghton} obtained a somewhat different result for the 
fermion Green function via a two dimensional bosonisation method.
Their result is equivalent to applying our previously discussed $N=0$
bosonisation technique to the half-filled Landau level problem.
As explained in Section III we do not believe this is a correct procedure.

We now turn to polarization bubble and vertices.
As in the previously considered spin liquid case, the only singularities occur
in the $2p_F$ vertices.
The leading $2p_F$ vertex correction, Fig.~\ref{F5}a, is given after summing
over parallel momenta by
\begin{equation}
\Gamma_{2p_F}^{(1)} = v_F \sum_{\epsilon} \int 
	\frac{(dk)}
	{\frac{p_0|\epsilon|}{2\pi|k|} + \frac{e^2}{8\pi \hat{\epsilon}} |k|}
	\frac{1}{\frac{2 \hat{\epsilon} v_F}{\pi e^2} |\epsilon| 
	\ln \epsilon +\frac{v_F}{p_0} k^2 }
\end{equation}
Again, the leading contribution to the integral over $k_\perp$ is a logarithm 
coming from the region $\epsilon < v_F k_\perp^2/p_0 < \epsilon \ln \epsilon$.
Performing this integral and evaluating the sum over frequencies we get
\begin{equation}
\Gamma_{2p_F}^{(1)} = \frac{1}{2} \ln^2 \left[ \ln\left( \frac{\epsilon_F}
	{max(T,\omega,v_F (Q-2p_F)^2/p_0)} \right)\right]
\end{equation}

Although it is of only academic interest, we note that  the higher order corrections may be summed to obtain the leading singular
behavior.
As in the case of the self-energy, crossed graphs are less singular than ladder
ones. 
As in Section II, the sum of the ladder graphs exponentiates, leading to 
\begin{equation}
\Gamma_{2p_F} = \exp\left[\frac{1}{2} \ln^2 \left[\ln\left( \frac{\epsilon_F}{T} 
	\right) \right] \right]
\label{Gamma_2p_F_hf}
\end{equation}

This weak singularity implies that the polarizability is not singular, but the
leading frequency and momentum dependence is weakly singular.

\section{Scaling}

In this section we recover some of the results obtained in previous
sections via a scaling analysis.
Our principal result concerns the properties of the effective action $S_{eff}$
of dressed fermions, $\Psi$, coupled to a gauge field ${\bf a}$ in $d$ spatial
dimensions:
\end{multicols}
\begin{eqnarray}
S_{eff} &=&\int (d\omega d^dk) \Psi_{\omega,k}^{\dagger}[|\omega|^{d/(2+x)} -
v_F k_\parallel + \frac{v_F}{2p_0} k_\perp^2] \Psi_{\omega,k}
+ \int (d\omega d^dq) (|q|^{1+x} + \frac{p_0|\omega|}{|q|}) |a(\omega,q)|^2 
\nonumber \\
&+& g \int (d\omega_1 d^dk_1)(d\omega_2 d^dk_2) \Psi_{\omega_1,k_1}^{\dagger}
	\Psi_{\omega_2,k_2} [{\bf v}_F(k_1) + {\bf v}_F(k_2)]
	({\bf a}(\omega_1 \!-\! \omega_2,k_1 \!-\! k_2)+h.c)
\label{S_eff}
\end{eqnarray}
\begin{multicols}{2}

We find that the fermion-gauge-field interaction $g$ is irrelevant for $d>2$ 
and marginal in $d=2$.
Further, for $x>0$ in $d=2$ the marginality of the interaction leads to
logarithms only in the $2p_F$ response functions. 

Note that $S_{eff}$ involves dressed fermions with one-loop self-energy
$\Sigma=|\omega|^{d/(2+x)}$ rather than the linear $\omega$ dependence
expected for unrenormalized fermions.
As shown by Nayak and Wilczek \cite{nayakwilczek}, if the linear $\omega$ 
dependence is used in
$S_{eff}$, then the fermion-gauge-field interaction $g$ is relevant for
$d<2+x$, and is in particular relevant in $d=2$ for $x>0$.
We argue that the strong-coupling fixed point to which the Nayak-Wilczek 
scaling flows is simply the $S_{eff}$ we have written above.
The argument has two steps.
The first is the known result \cite{lee} that the first order correction to
the fermion propagator from the gauge field interaction is of the form
$|\omega|^{d/(2+x)}$.
The second step is that further corrections do not change the form
given by the first order correction.
We have shown this in previous sections by explicit solutions of the model in
two limits.
In this section we give a scaling argument leading to the same conclusion.

We first explain our choice of notation in more detail;
it comes from the fact, seen in the calculations of the previous sections,
that a gauge fluctuation of momentum $\bf q$ couples primarily to fermions in
a patch of the fermi surface where the fermion velocity $\bf v_F$ is perpendicular to the 
direction of $\bf q$.
Therefore, in the effective action we have written the momentum dependence of
the fermion fields using local coordinates defined in a patch centered on the
point (in $d=2$) or strip (in $d=3$) of the fermi surface where $\bf v_F$ is perpendicular to $\bf q$.
In this patch the gauge-field-fermion interaction is simplified because the
transverse component of the gauge field is almost parallel to $\bf v_F$, so we
may replace the cosine of the angle between the gauge field $\bf a$ and $\bf 
v_F$ by unity.

To see that this construction is reasonable, note that from the gauge field
propagator in $S_{eff}$ we learn that at frequency $\omega$ the important
momentum scale is $|\omega|^{1/(2+x)}$.
>From the fermion propagator we learn that the important momentum scale in the
direction perpendicular to the fermi surface is $|\omega|^{d/((2+x)}$;
for $d>1$ this is always much less than the scale defined from the gauge
propagator, so that the momentum transferred from the gauge field to the
fermion is essentially perpendicular to the fermi velocity, and the patch
construction is well defined.
Also from the fermion propagator we see that the $k_\perp$ scale is
$|\omega|^{\frac{d}{2(2+x)}}$.
Thus in $d>2$ the dependence of the fermion propagator on $k_\perp$ is
essential.
In $d=2$, the momentum scale derived from the gauge field and from the fermion
propagator are the same, and the importance of the curvature term
$k_\perp^2/(2p_0)$ is determined by a dimensionless parameter (e.g. $N$).
We see that in these arguments the curvature of the Fermi surface (specified
by $p_0^{-1}$) is essential.
We shall show below that  $p_0^{-1}$ changes under scaling, so one must
interpret it as a charge in the
renormalization group equations.

We now discuss the ``tree-level'' scaling procedure.
The theory has three coordinates: frequency, $k_\parallel$, and $q$ (which we
have shown is the same as $k_\perp$).
All three scale differently.
We choose the scaling of $k_\parallel$ following Shankar \cite{shankar}: 
that is, we imagine
integrating out fermions in a shell given by $\Lambda/b<\epsilon_k<\Lambda$
about the Fermi surface and then rescaling the momentum perpendicular to the
fermi surface to restore the upper cut off.
We then choose the scaling of frequency to keep the $|\omega|^{d/(2+x)}$ term
in the fermion action invariant, and then choose the scaling of $q$ (which is the
same as that of $k_\perp$) to
keep the gauge field propagator invariant.
Finally, we choose the scaling of the fields to compensate for the scaling of
the coordinates and integrals, so that the quadratic terms remain invariant.
Note that we must interpret $\int d^dq$ in the gauge field term or the fermion
field term as $\int dq_\parallel d^{d-1}q_\perp$.
This implies
\[
\Psi \rightarrow \Psi b^{\frac{3d+1+x}{2d}} 
\]
\[
a    \rightarrow a b^{\frac{d+1+x}{d}}
\]
Combining all factors we get the following tree-level scaling equations for
the charges $1/p_0$ and $g$:
\begin{eqnarray*}
\frac{d p_0^{-1}}{d \ln b} &=& \left(1- \frac{2}{d}\right) p_0^{-1} \\
\frac{d g}{d \ln b}      &=& 0
\end{eqnarray*}

Therefore, the gauge-field fermion coupling is marginal for $d \geq 2$ but in
$d>2$ the effective curvature of the fermi surface grows.
For large curvature the usual arguments leading to the Migdal theorem \cite{migdal} imply that
the crossed graphs may be neglected, so we may restrict ourselves to the
leading order of perturbation theory, which gives the self energy 
$|\omega|^{\frac{d}{2+x}}$.

Alternatively, one may consider a scaling procedure which preserves the form
of the fermion propagator.
In this case one must scale $k_\perp$ as $b^{-1/2}$ and in $d>2$ both the
coefficient of the $q^{1+x}$ term in the gauge propagator and
gauge-field-fermion coupling $g$ scale to zero (indeed the tree-level scaling
equation for $g$ becomes $\frac{dg}{d \ln b} = \frac{2-d}{4}$, so that a
manifestly weak coupling fixed point is obtained).

In $d=2$, however, all charges are marginal for all $x>0$ and further analysis
beyond tree level is needed to determine which physical quantities are
renormalized. 
For our purposes the most efficient method of deriving the one-loop
renormalization group equations is to use the technique of differentiating the
one-loop diagrams with respect to the upper cut off.
The calculations presented in the previous sections can be carried over
directly to show that the only quantities which are renormalized are the $2p_F$
vertex $\Gamma_{2p_F}$ and polarizability $\Pi(2p_F)$.
In particular, neither $p_0$ nor $g$ scales in $d=2$.
Rewriting  Eq. (\ref{delta_Gamma}) in the notations of this section (here we
normalize to $k_\parallel$ and in the previous section we normalized to
frequency) we find
\begin{equation}
\frac{d \Gamma_{2p_F}}{d \ln b} = \frac{2+x}{2} \sigma \Gamma_{2p_F}
\end{equation}
where $\sigma$ is a number which depends on the fixed point values of $p_0$
and $g$.
Our results of the previous sections may be viewed as calculations of the
fixed pont values of $p_0$ and $g$ in the large and small $N$ limits.

Although there are no logarithmic corrections to the fermion propagator, the
finite renormalizations generated by marginally irrelevant operators do mean
that the fermion propagator $G^{-1}(\epsilon,p_\parallel,p_\perp)$ is not
precisely given by the form $|\epsilon|^{\frac{2}{2+x}} - v_F (p_\parallel +
p_\perp^2/(2p_0))$ written in Eq. (\ref{S_eff}) when
$|\epsilon|^{\frac{2}{2+x}}$, $v_F p_\parallel$ and $p_\perp^2/(2p_0)$ are of
the same order, although the limits when one argument is much larger than the
others are correctly given. 

Finally, we consider $d=2$, $x=0$.
Here at tree level we would conclude that for the action with inverse
fermion propagator
$(\omega - v_F k_\parallel + \frac{v_F}{2p_0} k_\perp^2)$ and for inverse 
gauge propagator
$(|q| + \frac{|\omega|}{|q|})$ the fermion-gauge-field coupling is marginal.
However, caution should be excercised in deriving renormalization group
equations beyond tree level, because in the two-loop calculations presented in
the previous section no terms of order $(\ln \Lambda)^2$ were found so that
the logarithms found in one-loop order do not sum to powers.
Instead, the calculations presented in section IV show that the asymptotic form of the fermion propagator is 
$i \omega \ln |\omega| - v_F k_\parallel - \frac{v_F}{2p_0} k_\perp^2$,
and the $2p_F$ vertices are extremely weakly singular 
($\sim \exp \frac{1}{2} \ln^2\ln \omega$).

\section{Conclusion}

We have presented a discussion of the low energy properties of a system of
fermions in spatial dimension $d$ coupled via a singular gauge interaction
with propagator $D(\omega,q)=(|\omega|/|q| + |q|^{1+x})^{-1}$.
We found that the fermion lifetime scales as $|\omega|^{\frac{d}{2+x}}$ (in
$3\geq d \geq 2$, $x>0$) and that in $d=2$ the $2p_F$ fermion polarizability
$\Pi(\omega,q)$ was non-analytic and possibly divergent as $Q\rightarrow 2p_F$
and $\omega \rightarrow 0$.
Whether or not the susceptibility is divergent depends on the value of an
exponent, $\sigma$, which we could calculate only in certain unphysical
limits.
In the spin liquid case $x=1$ extrapolation of our calculated $\sigma$ to the
physical limit of spin degeneracy $N=2$ from two sides yielded estimates for $\sigma$ bracketing the
critical value $\sigma_c=1/3$ above which $\Pi$ diverges.
In the $\nu=1/2$ case one must distinguish between screened and unscreened
Coulomb interactions.
In the unscreened case, $x=0$, the self energy is $\omega \ln \omega$ while
the nonanalyticity in the $2p_F$ vertex is very weak: 
$\exp\left( \frac{1}{2} \ln^2[ \ln[\omega]] \right)$ and the polarizibility does not diverge.
In the screened case the results for $x=1$ apply with spin degeneracy $N=1$ and our estimates suggest
that the $2p_F$ polarizability diverges.

There is a simple physical interpretation for the nonanalyticities at $2p_F$:
a moving fermion emits a gauge field which relaxes so slowly that if at a
later time the fermion is scattered backwards it meets the gauge field again
and is able to lower its energy.
It is remarkable that this physics can lead to an actual divergence of the
$2p_F$ susceptibility if the fermion-gauge-field interaction is strong enough.
The form of the divergence is given in Eq. (\ref{Pi_2D},\ref{Pi_1D}) and is
controlled by an exponent $\sigma$ which can a priori take any value.
However, we note that if $\sigma \geq 7/6$, 
then $\sum_{\omega,q} \Pi(\omega,q)$ is infrared divergent.
Such a divergence is not possible; for example in a magnetic system this would
imply that the expectation value of the square of the local spin density
$\langle S_i^2 \rangle$ diverges.
Therefore we believe that for $\sigma \geq 7/6$ some other physics beyond the
scope of our calculations must intervene to cut off the divergence.
One mechanism for this feedback can be seen in the spin-fluctuation 
contribution to the electron self-energy.
For $\sigma \geq 7/6$ this diverges, implying a smearing of the Fermi surface
which would suppress the Fermi surface singularities we have found.
However, for $7/6 > \sigma > 1/3$ we believe this critical phase is stable.

In order to understand the physical properties of the critical phase, consider
first a translation-invariant electron gas (as is realized in the half-filled
Landau level).
Then Eq. (\ref{Pi_1D}) would predict that the susceptibility diverges as $T
\rightarrow 0$ on a ring of radius $2p_F$.
For fermions on a lattice, the situation is more complicated for  reasons
very similar to those analysed by Littlewood et al \cite{littlewood} in a 
study of $2p_F$ singularities in a marginal Fermi liquid picture.
First, intead of a circle of radius $2p_F$ one obtains one or more curves
traced out by the vectors $\bf Q$ connecting points with parallel tangents.
Second, the amplitude (but not the exponent) of the divergent term in $\chi_{2p_F}$ varies around the curve due to
the variation of $v_F$ and $p_0$ around the Fermi surface.
Third, one obtains additional families of curves on which $\chi$ diverges. 
These are generated by $\bf Q + G$ where 
$\bf G$ is any vector of the reciprocal lattice. As a result one gets
additional peaking when members of different families intersect.
The result, for band structures appropriate to high-$T_c$ superconductors will
be a susceptibility strongly peaked at particular points in $q$-space which
might be qualitatively consistent with neutron data for $La_{2-x} Sr_x Cu
O_4$ \cite{littlewood}.  In addition,
the divergent spin fluctuations imply that the $Cu$ NMR rate $1/T_1 T \propto
1/T^{2\sigma-1/3}$, so the $1/T_1 T$ diverges as $T\rightarrow 0$ even at the
borderline value $\sigma=1/3$.
The value $\sigma=2/3$ would lead to $1/T_1T \propto 1/T$ consistent with $Cu$
NMR experiment on high-$T_c$ superconductors.
Of course, if these wavevectors where $\chi''(\omega,q)$ is maximal are too
far from the commensurate wavevector ($\pi,\pi$), the oxygen $1/T_1T$ will
also diverge, in disagreement with experiment \cite{millis}.

In the half-filled Landau level case with screened Coulomb interaction the
divergence in the $2p_F$ susceptibility could in principle be observed in 
sound propagation experiments in which the phonon wavevector is tuned to
$2p_F$. 
The divergence should lead to a large damping of the phonon which increases as
$T$ is decreased.
The effect should be observable for temperatures and phonon frequencies less
than a scale $\omega_0$ which we calculate from Eqs.
(\ref{Sigma^(1)},\ref{omega_0},\ref{tildeD}).
We rewrite the expression for $\omega_0$ in terms of the Fermi energy
$\epsilon_F=\hbar^2 p_F^2/(2m)$, Coulomb parameter $E_c=\frac{e^2 n^{1/2}}
{\hat{\epsilon}}$ and the screening length $\kappa^{-1}$, obtaining
\begin{equation}
\hbar \omega_0 \approx 0.15 \frac{\epsilon_F^3 \kappa^2}{n E_c^2}
\label{omega_0_est}
\end{equation}
Assuming typical numbers for $Ga AL As$ inversion layers $m=0.07 m_e$, 
$\hat{\epsilon}=13$ and $n=10^{11} \; cm^{-2}$ we have $E_F \sim 50 \;K$ and
$E_c \sim 40\; K$ so $\hbar \omega_0\;[K] \approx 10 \frac{\kappa^2}{n}$.
Thus if the screening length is not too much greater than the interparticle
spacing, the effect should be observable. 
 
\end{multicols}
\begin{appendix}
\begin{multicols}{2}

\section{Vertex at low momentum transfer}

Here we use the Ward identity to derive the exact form of the renormalized 
fermion-gauge-field vertex at low momentum transfer $q_\perp$ and $q_\parallel$
(the exact conditions under which this form applies will be obtained below).
In this limit the renormalized vertex becomes singular, and our goal is to find
the form of this singularity.
The expression that we shall find is correct at any $N$ for sufficiently small 
transferred momenta $q_\perp,q_\parallel$.
The value of N determines only the range
of $q_\perp$, $q_\parallel$ over which the expression obtained in the limit
of very low momenta remains valid.

The fundamental Ward identity was given in Eq. (\ref{Ward}); we
repeat it here for convenience.
\end{multicols}
\[
i \omega \Gamma^{0}_{\epsilon,p} (\omega,q) - 
q_\parallel \Gamma^{\parallel}_{\epsilon,p} (\omega,q) - 
q_\perp \Gamma^{\perp}_{\epsilon,p} (\omega,q) = 
	G^{-1}(\epsilon+\omega/2,p+q/2) - G^{-1}(\epsilon-\omega/2,p-q/2)
\]
\begin{multicols}{2}
Here $\Gamma^0$ is the density vertex and $\Gamma^{\parallel}$ and 
$\Gamma^{\perp}$ are the components of the current vertex parallel and
perpendicular to $\bf v_F (\bf p)$.
We wish to obtain from this an equation relating $\Gamma^0$ to the fermion
Green function.  As noted previously, in the present model at small $q,\omega$
the current vertex $\Gamma^{\parallel}$ is related to the density vertex
$\Gamma^{0}$ by $\Gamma^{\parallel}=v_F\Gamma^{0}$.  We now derive the relation
between $\Gamma^{\perp}$ and $\Gamma^0$.
Consider a  high order diagram for $\Gamma^{\perp}$ of the type  shown in
Fig.~\ref{F9}a, in which ${\em one}$ of the gauge field lines connecting
one external fermion leg to the other is isolated, i.e. not
crossed by any other gauge field line connecting one external fermion
leg to another.
The analytical expression has the general form
\begin{equation}
\Delta \Gamma^\perp=\frac{v_F}{p_0} \int \prod_j (d^2 k_j) \sum_{i=1}^{i=n} 
	k_{\perp i} \prod_j D(k_j) \prod G
\label{Delta_Gamma}
\end{equation}
where index $i$ runs over $n$ values corresponding to the gauge field
lines connecting
different legs and we did not explicitly write the arguments of the
fermion  Green functions $G$.  
Label the momentum of an "uncrossed" gauge field line by ${\bf k}_a$,
and pick out the term in the sum proportional to $k_{\perp a}$.
We show below that all other terms in the sum are small.

In the limit $q_\perp \rightarrow 0$ the fermion Green functions in
diagrams such as Fig.~\ref{F9}a
depend only on the combination $(k_{\parallel a} +
k_{\perp a}^2/2p_0)$, so their dependence on $k_{\perp a}$ can be completely
eliminated by the shift $k_\parallel \rightarrow k_\parallel - k_\perp^2/2p_0$.
(Recall that the $k_{\parallel}$ dependence of D is negligible always).
After this transformation the only remaining dependence on $k_{\perp a}$ is in 
$D(k_{\perp a})$.
The remaining integration over $k_\perp$ is straightforward:
\begin{equation}
\int D(k_{\perp a}) k_{\perp a} dk_{\perp a} =0
\label{integral}
\end{equation}
This integral converges poorly at large $k_\perp$, because the 
integrand decreases as $\sim 1/k_\perp$ at large $k_\perp$, but is zero at
$q_\perp=0$, because the integrand is an odd function of $k_\perp$.
At any $q_\perp \neq 0$ the dependence of the Green fuctions on the
momentum $k_\perp$ can no longer be neglected.
Since Green functions depend only on the product $q_\perp k_\perp$,
their $k_\perp$ dependence becomes significant only at large 
$k_\perp \sim \Sigma(\epsilon)/q_\perp$.
At smaller $k_\perp$ the dependence of the Green function on $k_\perp$ can be
neglected and the contributions from positive and negative $k_\perp$ cancel
each other. Because the main contribution to this diagram comes from large
$k_{\perp a} \sim \Sigma(\epsilon)/q_\perp$, the  Migdal arguments of
Sections II and III show that at large $k_\perp$ all diagrams in which other
lines cross the line with large momentum transfer are small. 

Now consider any arbitrary diagram for $\Gamma^{\perp}$.  The corresponding analytical expression will be of the form shown in eq. (\ref{Delta_Gamma}).
Pick out one term in the $\sum_{i=1}^{i=n} k_{\perp i}$.  The diagram will be 
important only if the gauge field line carrying this momentum is "uncrossed"
in the sense discussed above.  
This justifies the assumption made above that the term in the sum
proportional to $k_{\perp a}$ corresponds to an ``uncrossed'' line.
Therefore, the diagrams which give the dominant contribution at small 
$q_\perp$ can be represented as the diagram shown in Fig.~\ref{F9}b.
Here the two blocks involve gauge field lines which cross each other
and the double wavy line represents $D(k_{\perp a}) k_{\perp a}$.
Since the integral over $k_\perp$ in the double wavy line is dominated by large
$k_\perp$, the frequency dependence of it can be neglected while the
dependence on $k_\parallel$ can be neglected always.  For
this reason, the outer
block is simply with the bare vertex $\Gamma^{0}$.
The inner block can be also expressed in terms of the vertex
$\Gamma^{0}$.  
After some manipulation we find:
\end{multicols}
\begin{eqnarray}
\Gamma^{\perp}_{\epsilon,{\bf p}}(\Omega,{\bf q}) &=& 
	\frac{N^{1/2} g^2 v_F^3}{p_0}
	\Gamma^{0}_{\epsilon,{\bf p}}(\Omega, {\bf q}) 
\label{Gamma^perp} \\
& \times &
	\int \Gamma^{0}_{\epsilon \!+\!\omega,{\bf p \!+\! k}}(\Omega, {\bf q})
	G(\epsilon \!+\! \omega \!+\! \Omega/2,p \!+\! k \!+\! q/2)
	G(\epsilon \!+\! \omega \!-\! \Omega/2,p \!+\! k \!-\! q/2)
	\frac{(d^2 k d\omega)}{k_\perp}
\nonumber
\end{eqnarray}
In this equation the vector character of the vertex is expressed via the
factor of $k_{\perp}^{-1}$ which may be positive or negative.
Together with Ward identity Eq. (\ref{Ward}), Eq. (\ref{Gamma^perp}) 
forms a closed system of equations for the vertex $\Gamma^{0}$. 
To solve it we introduce the notation 
\begin{equation}
\Gamma^{\perp}_{\epsilon,{\bf p}}(\omega,{\bf q}) = 
	v_F B(q_\parallel,q_\perp) 
		\Gamma^{0}_{\epsilon,{\bf p}}(\omega,{\bf q})
\label{notation}
\end{equation}
Here we have suppressed the dependence of
B on the frequency $\omega$, because $\omega$ is a dummy variable in the
analysis that follows.
We use the Ward identity to express $\Gamma^{0}$ in Eq. (\ref{Gamma^perp})
through $B(q_\parallel,q_\perp)$, finding:
\begin{equation}
B(q_\parallel,q_\perp)= \frac{v_F g^2 N^{1/2}}{p_0} 
	\int \frac{(d^2 k d \omega)}{k_\perp} 
	\frac{
	G(\epsilon \!+\! \omega \!+\! \Omega/2,p \!+\! k \!+\! q/2) -
	G(\epsilon \!+\! \omega \!-\! \Omega/2,p \!+\! k \!-\! q/2) }
	{i \Omega - v_F(q_\parallel + \frac{q_\perp k_\perp}{p_0}) 
	- q_\perp v_F B(q + \frac{q_\perp k_\perp}{p_0},q_\perp)}
\label{B0}
\end{equation}
Integrating over $k_\parallel$ and $\omega$ and scaling the $k_\perp$ variable
($k_\perp \rightarrow k_\perp p_0/q_\perp$) we find:
\begin{equation}
B(q_\parallel,q_\perp) = \alpha \; sgn q_\perp \;P\!\int \frac{dk}{2 \pi k}
\frac{i \tilde{\Omega}}{ i \tilde{\Omega} - (q_\parallel+k) - B(q_\parallel+k,
q_\perp) q_\perp}
\label{B1}
\end{equation}
\begin{multicols}{2}
\noindent
where $\tilde{\Omega}=\Omega/v_F$ and 
\begin{equation}
\alpha= \frac{v_F g^2N^{1/2}}{p_0}
\label{alpha}
\end{equation}
is a dimensionless parameter of the order of $N^{1/2}$.
It is convenient to consider positive and negative $\Omega$ separately. The
considerations are similar so we consider explicitly only $\Omega > 0$ here.
Equation (\ref{B1}) can be simplified if we assume that the 
denominator in it has poles only in the upper half plane.
We shall show this assumption is self-consistent.
In this case the integration contour can be closed in the lower half plane
and the integral equation simplifies to the algebraic equation
\begin{equation}
B(q_\parallel,q_\perp)=\frac{1}{2} \frac{\alpha \tilde{\Omega}}
	{i \tilde{\Omega} - (q_\parallel) - 
		q_\perp B(q_\parallel, q_\perp)}
\label{B2}
\end{equation}
Solving it we find
\begin{equation}
B(q_\parallel, q_\perp) = \frac{i\tilde{\Omega}-v_Fq_\parallel + 
	\sqrt{(i\tilde{\Omega} -v_F q_\parallel)^2 - 2\alpha |q_\perp| 
		|\tilde{\Omega}|}}
	{2 q_\perp}
\label{B3}
\end{equation}
Restoring the notations of Section III and using 
$v_F q_\parallel \gg \Omega$ we get the equation (\ref{B}) announced 
in Section III.

Combining the Ward identity (\ref{Ward}) with Eqs. (\ref{notation}) and 
(\ref{B}) we get the final expression for the vertex at low momentum transfer:
\begin{equation}
\Gamma^{(0)}_{\epsilon,{\bf p}}(\Omega,{\bf q}) = 
	2 \frac{
	G(\epsilon \!+\! \Omega/2,p \!+\! q/2) -
	G(\epsilon \!-\! \Omega/2,p \!-\! q/2)}
	{ v_F q_\parallel + \sqrt{(v_F q_\parallel)^2 - 2\alpha |\Omega|
	|q_\perp| -i\Omega v_Fq_\parallel}}
\label{Gamma^(2D)1}
\end{equation}

Thus, $\Gamma^{\perp}$ is substantially enhanced at low momenta $v_F
q_\parallel \ll |\omega_0^{1/3}| |\omega|^{2/3}$, $q_\perp < (v_F
q_\parallel)^2/(\alpha |\Omega|)$.
This range of momenta becomes small at $N \gg 1$ because $\alpha \sim
N^{1/2}$ and does not contribute much to the self-energy.

The essential ingredient in the derivation of (\ref{B}) and
(\ref{Gamma^(2D)1}) was assumption that the momentum $k_\perp$ is sufficiently
large so that crossing diagrams can be neglected. 
We also assumed that  $k_\perp \gg k_\omega$ which allowed us
to neglect the frequency dependence of the gauge field propagator.
Both these conditions are satisfied if $q_\perp \ll \frac{1}{N}
k_\Omega$.
In the limit $N \gg 1$ this condition limits drastically the range of 
momentum where (\ref{Gamma^(2D)1}) can be applied.

\section{Higher order diagrams in $1/N$}

The calculations of Appendix A show that the gauge-field-fermion vertex is
enhanced at very low momentum transfer as is evident from Eq. 
(\ref{Gamma^(2D)1}), 
This equation, however, was derived under the assumption that 
$q_\perp \ll \frac{1}{N} k_\omega$.
At larger momentum transfer $q_\perp$ the corrections to the bare vertex are
small, Eq. (\ref{Gamma^(2D)1}) is not valid, instead, the leading corrections
are given by the first crossing diagram shown in Fig.~\ref{F3}.
The straightforward calculation gives Eq. (\ref{Gamma^(1R)}).
The equation (\ref{Gamma^(1R)}) crosses over to the equation 
(\ref{Gamma^(2D)1}) at $q_\perp \sim \frac{1}{N} k_\omega$.
So, at large $N$ the momentum range where the whole series of diagrams
should be summed is small in $1/N$, morreover, this momentum range
turns out to be so small that it does not contribute even to
subleading order in $1/N$ for most quantities.
For instance, the contribution of this range to the self energy is of the
order of $\frac{1}{N^2}$, whereas the leading term which comes from larger 
momenta is of the order of $\left( \frac{\ln[N]}{N} \right)^2$ (Eq.
(\ref{Sigma^(2)_2})).
Thus, in order to obtain the subleading terms of the order of $\frac{\ln^2[N]}
{N^2} $ it is sufficient to keep only the first crossing diagrams in photon
propagator.
However, in order to obtain all terms of the order of $\frac{1}{N^2}$ one
needs to use the Eq. (\ref{Gamma^(2D)1}) and the crossover formulas (which
we did not derive) in the range $q_\perp \sim \frac{1}{N} k_\omega$.

Similarly, in the calculation of the exponent of the $2p_F$ vertex, the
enhancement of the gauge-field-fermion vertex at 
$q_\perp \lesssim \frac{1}{N} k_\omega$
leads to  corrections of the order of $\frac{1}{N^2}$ to the exponent.
As we shall see below, the leading terms are larger by factors of $\ln[N]$,
so in the subleading term we again can keep only the simplest crossing
diagrams shown in Fig.~\ref{F5}.
Only the diagrams shown in Fig.~\ref{F5}b and Fig.~\ref{F5}d have
contributions which contain powers of $\ln[N]$.
Consider the diagram shown in Fig.~\ref{F5}b.  Its analytical expression
reads
\end{multicols}
\begin{equation}
\delta_1 \Gamma_{2p_F}(\omega) = v_F^4 \int D(\eta,k)  D(\Omega,q)
		G(\eta,k)G(\Omega+\eta,q+k)G(\Omega,q)G(\Omega+\omega,-q)
		 (d^2q d^2k d\Omega d\eta)
\end{equation}
The contribution containing logarithms of $N$ comes from the momentum range
$k_\eta > |k_\perp| > |q_\perp| > \frac{1}{\sqrt{N}}  k_\Omega$.
In this range the self energy parts of the Green functions can be neglected.
We integrate over parallel components of momenta $k_\parallel$ and
$q_\parallel$  and symmetrize the resulting expression obtaining: 
\begin{equation}
\delta_1 \Gamma_{2p_F}(\omega) = \frac{p_0^2}{2 \pi^4}  \int_\omega^\infty
d\Omega \int_0^\Omega d\eta \int_0^\infty \frac{dk_\perp dq_\perp}
{k_\perp^2 q_\perp^2 - q_\perp^4} 
\end{equation}
Evaluating the remaining integrals with logarithmic accuracy we get
\begin{equation}
\delta_1 \Gamma_{2p_F}(\omega) = \frac{1}{8 \pi^2} \frac{\ln^3[N]}{N^2}  
				\ln \left( \frac{1}{\omega} \right)
\end{equation}
Evaluation of the diagram shown in Fig.~\ref{F5}d is very similar 
It has analytical expression
\begin{eqnarray}
\delta_2 \Gamma_{2p_F}(\omega) &=& v_F^4 \int D(\eta,k)  D(\Omega,q)
\\
&\times& 
		G(\Omega,k)G(\Omega+\eta,q+k)G(\Omega+\eta+\omega,-(q+k))
		G(\Omega+\omega,-q) (d^2q d^2k d\Omega d\eta)
\nonumber
\end{eqnarray}

We integrate over parallel components of momenta, obtaining after
symmetrization:
\begin{eqnarray}
\delta_2 \Gamma_{2p_F}(\omega) = \frac{p_0^2}{2 \pi^4}  \int_\omega^\infty
d\Omega \int d\eta \int_0^\infty &&
	\frac{\beta |\Omega|^{2/3}|\eta +  \Omega|^{2/3} -
			q_\perp^2(k_\perp+q_\perp)^2 
			sgn [\Omega (\Omega+\eta)]}	 
{(\beta |\Omega|^{4/3} + q^4)(\beta |\Omega+\eta|^4 + (q_\perp + k_\perp)^4)}
\nonumber \\
&\times& D(\Omega,q_\perp) D(\eta,k_\perp) dk_\perp dq_\perp
\end{eqnarray} 
\begin{multicols}{2}
\noindent
Here $\beta=4 \left(\frac{p_0}{v_F}\right)^2 \omega_0^{4/3}$.
The first term in the numerator of this integral is logarithmically divergent; 
the main contribution to the integral comes from the frequency range 
$\eta > \Omega > \omega$ and results in   
a $\frac{1}{2N^2}\ln^2(1/\omega)$ contribution which one expects from 
general renormalization group arguments.
The second term in the numerator has no contribution from this frequency range
due to $sgn [\Omega (\Omega+\eta)]$, so it results in only one power of 
$\ln(1/\omega)$, instead it contains $\ln[N]$ coming from the momentum range
$k_\eta > |k_\perp| > |q_\perp| > \frac{1}{\sqrt{N}}  k_\Omega$.
In this momentum range we neglect the self energy parts of the Green functions
and perform integrals with logarithmic accuracy obtaining:
\begin{equation}
\delta_2 \Gamma_{2p_F}(\omega) = \frac{1}{4 \pi^2} \frac{\ln^3[N]}{N^2}  
				\ln \left( \frac{1}{\omega} \right)
\end{equation}
Adding the contributions from the diagrams in Fig.~\ref{F5}b (which come with a factor
of two) and Fig.~\ref{F5}d
we get Eq. (\ref{delta_Gamma}).

\section{Exponent of the $2p_F$ vertex in the small $N$ limit}

\begin{figure}
\centerline{\epsfxsize=6cm \epsfbox{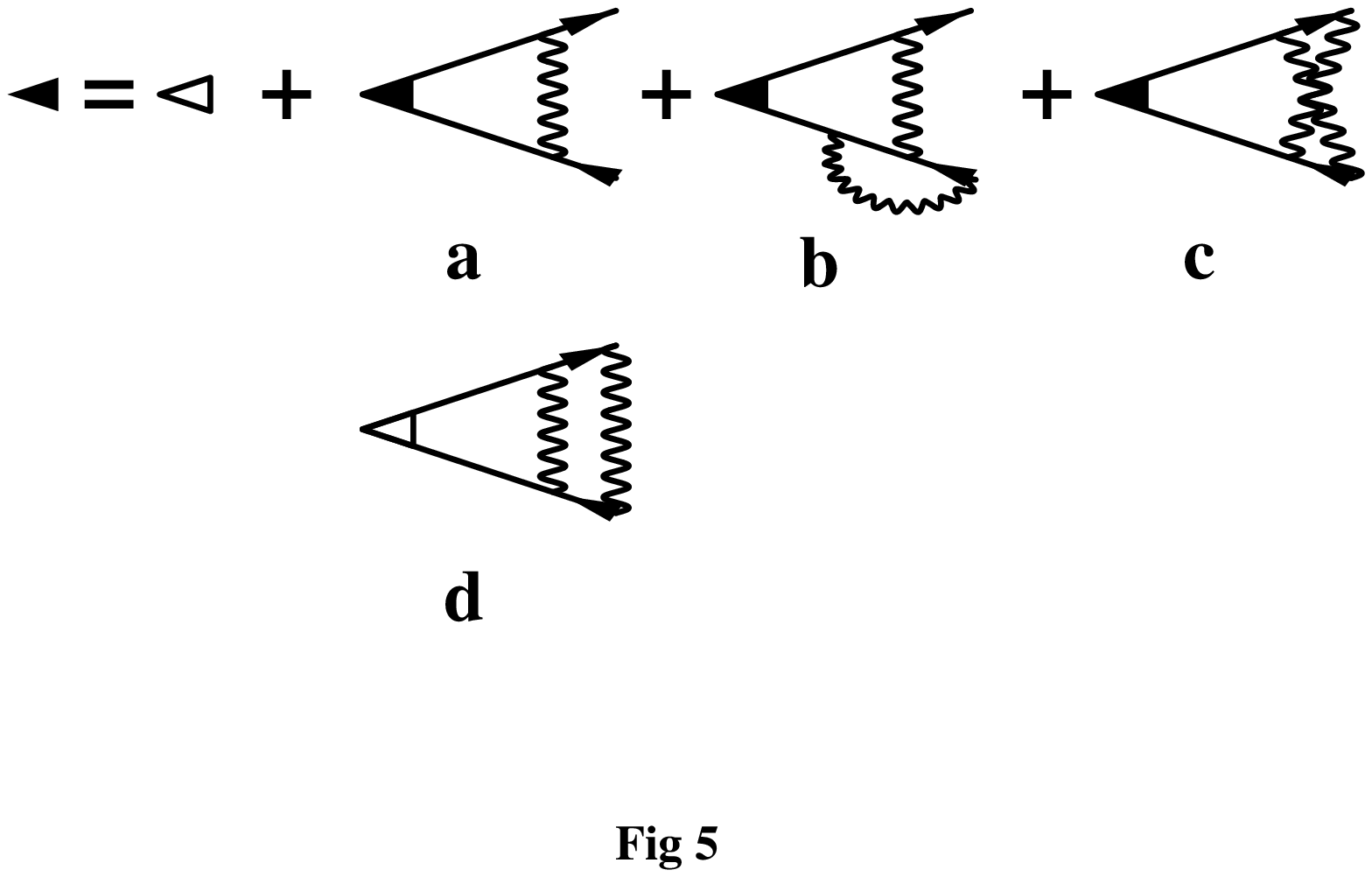}}
\caption{Ladder sum giving renormalization of fermion $2p_F$ vertex
$\Gamma_{2p_f}$ (shaded triangle). 
a. Leading order in $ln[N]/N$ b. and c. Subleading order in $1/N$.
d. The leading order in $(ln[N]/N)^2$.
The heavy dot indicates the bare $2p_F$ vertex, 
the wavy line denotes the gauge field propagator (2) and
the solid line the fermion propagator.}
\label{F5}
\end{figure}

\begin{figure}
\centerline{\epsfxsize=6cm \epsfbox{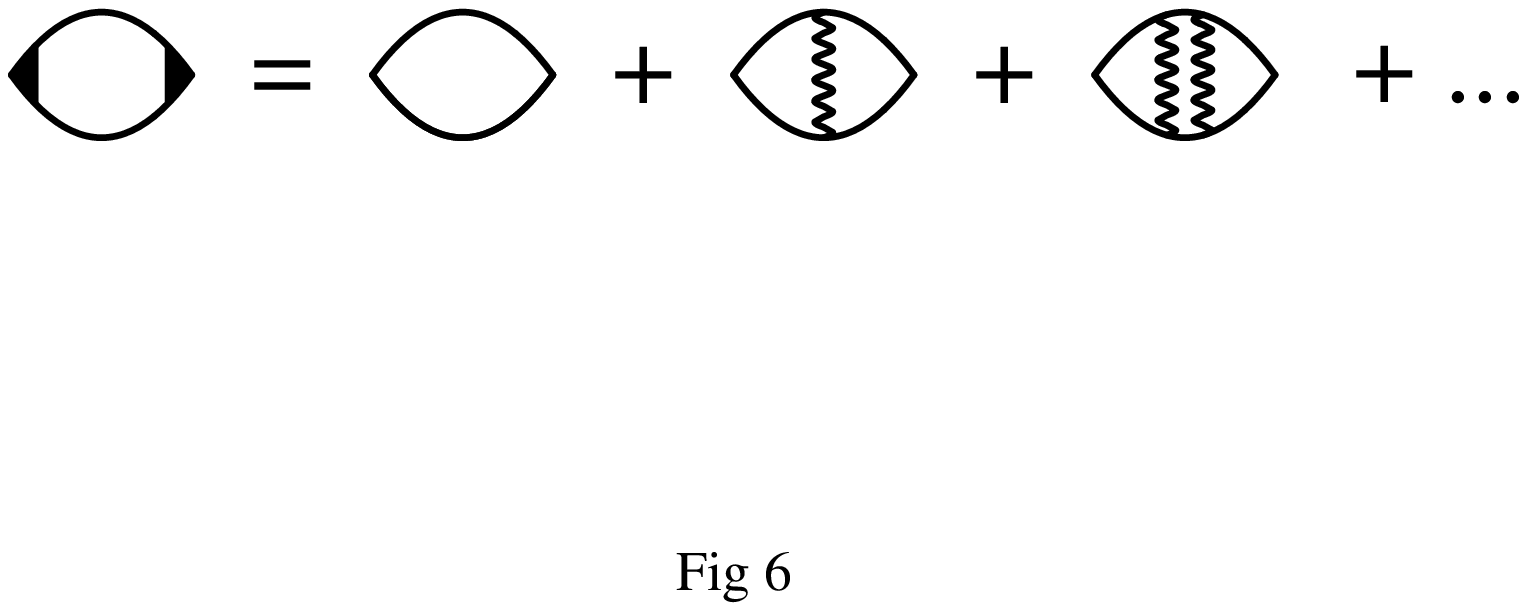}}
\caption{Ladder sums giving renormalization of fermion $2p_F$ polarizability.
Notation is the same as in Fig. 5}
\label{F6} 
\end{figure}

\begin{figure}
\centerline{\epsfxsize=6cm \epsfbox{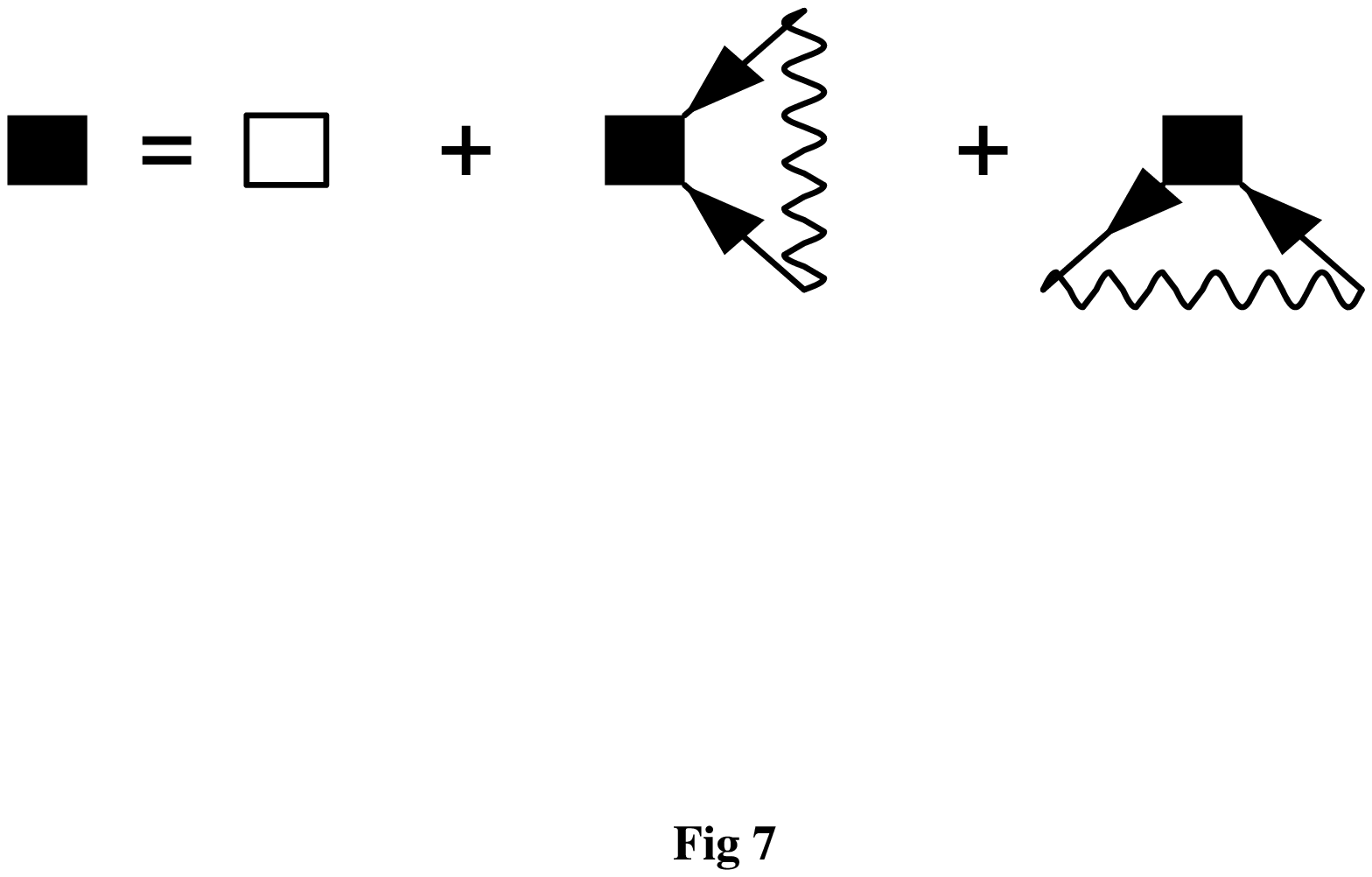}}
\caption{Diagrams leading to renormalization group equation for short range
fermion vertex $W$ (heavy square).
The wavy line denotes the gauge field propagator (2) and
the solid line the fermion propagator.}
\label{F7} 
\end{figure}

\begin{figure}
\centerline{\epsfxsize=6cm \epsfbox{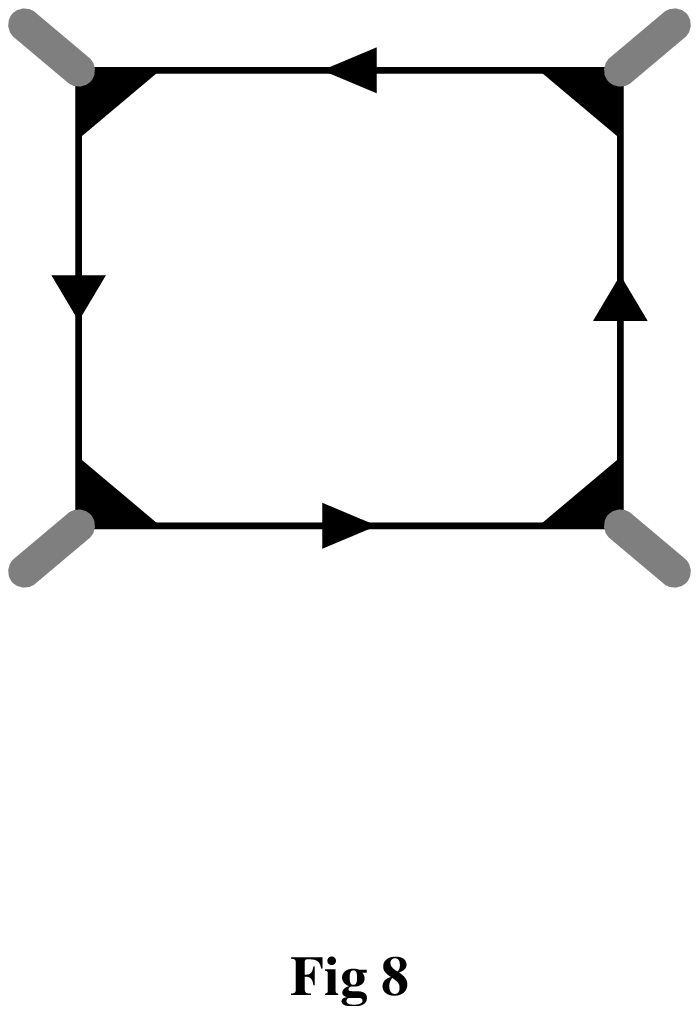}}
\caption{Diagram leading to a singular interaction between spin waves.
Solid lines denote fermion propagators, shaded triangles $\Gamma_{2p_F}$
vertex.}
\label{F8} 
\end{figure}

\begin{figure}
\centerline{\epsfxsize=6cm \epsfbox{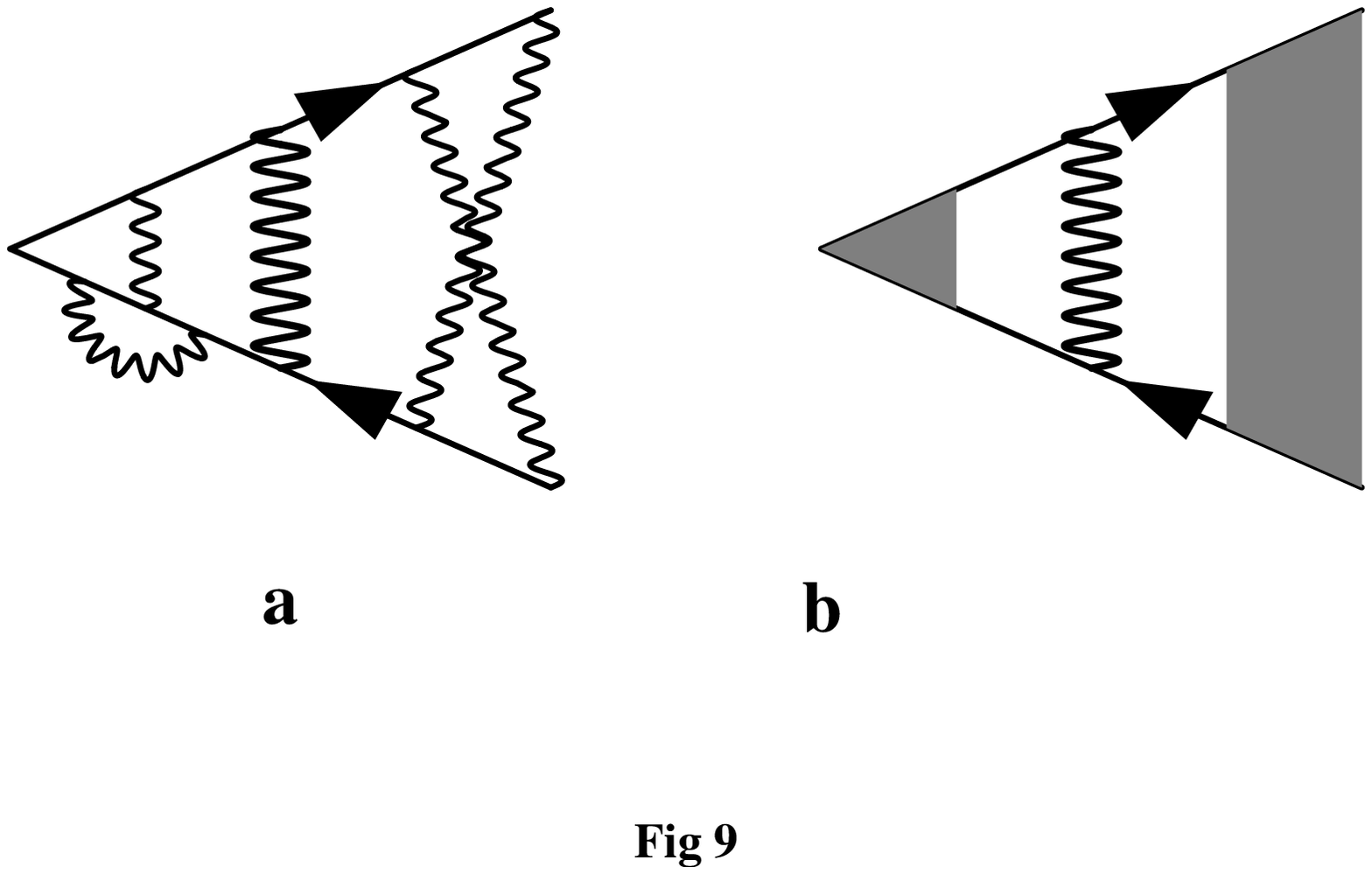}}
\caption{a. Typical high order diagram for $\Gamma^{\perp}$. b.
Sum of these diagrams. Broad wavy line represents propagator
$k_{\perp}D(\omega,k_\perp)$ with large momentum transfer in
perpendicular direction.}
\label{F9}
\end{figure}

In order to find the exponent of the $2p_F$ vertex in the small $N$ limit
we evaluate the first correction to the $2p_F$ vertex shown in Fig.~\ref{F5}a
using the exact gauge-field-fermion vertices and then exponentiate the result.
This prescription is known to work in 1D Lutinger model and it gives the
leading terms of the $1/N$ expansion.
The analytical expression for Fig.~\ref{F5}a is
\end{multicols}
\begin{equation}
\delta \Gamma_{2p_F}(\omega) = -v_F^2 \int G(\eta,q) G(\eta+\omega,-q)
	D(\eta,q) \Gamma_{\eta,q}(\frac{\eta}{2},\frac{q}{2})
	\Gamma_{-\eta,-q}(\frac{\eta}{2},
\frac{q}{2}) (d^2 q d\eta)
\end{equation}
This expression simplifies if the external frequency of the
fermion is zero and its momentum is on the Fermi surface because in this case
$G^{-1}(\epsilon,p) = 0$ in Eq. (\ref{Gamma^(0)}) for the vertex:
\begin{equation}
\delta \Gamma_{2p_F}(\omega) = \frac{1}{\pi^3} \int_\omega^\infty d\eta
	\int_0^\infty \frac{ 4 v_F^2 dq_\perp dq_\parallel}
	{\left| v q_\parallel + \sqrt{(v_F q_\parallel)^2 - 2
	\alpha \eta  q_\perp} \right|^2}
		D(\eta,q_\perp)
\end{equation}
\begin{multicols}{2}
Here we replaced the exact dependence on the external frequency $\omega$ by an
approximate cut off which is sufficient for logarithmic accuracy.
Evaluating this integral we get
\begin{equation}
\delta \Gamma_{2p_F}(\omega) = \sigma \ln\left( \frac{1}{\omega} \right)  
\end{equation}
where $\sigma$ is given by (\ref{sigma_2}).

\end{multicols}
\end{appendix}
\begin{multicols}{2}

\end{multicols}

\end{document}